\documentclass[11pt,a4paper]{article}
\pdfoutput=1
\usepackage[utf8]{inputenc}
\usepackage[english]{babel}
\usepackage{amsmath}
\usepackage{amsthm}
\numberwithin{equation}{section}
\usepackage{amsfonts}
\usepackage{mathtools}
\usepackage{mathrsfs}
\usepackage{mathpazo}
\usepackage{multirow}
\usepackage{amssymb}
\usepackage{graphicx}
\usepackage{ifpdf}
\usepackage{braket}
\usepackage{tikz-cd}
\usepackage{changes}
\newcommand{\changefont}[3]{
\fontfamily{#1} \fontseries{#2} \fontshape{#3} \selectfont}
\changefont{ppl}{m}{n}

\newcommand{\barf}{\overline{f}}

\usepackage[left=2.50cm, right=2.50cm, top=2.50cm, bottom=3cm]{geometry}

\theoremstyle{plain}

\theoremstyle{definition}

\theoremstyle{remark}

\usepackage[bookmarks=true,colorlinks=true,linkcolor=black,citecolor=black,urlcolor=black,bookmarksnumbered]{hyperref}
\usepackage{cite}

\begin{document}

\setcounter{equation}{0}
\setcounter{footnote}{0}
\setcounter{section}{0}

\thispagestyle{empty}

\begin{flushright} 
\texttt{LMU-ASC 11/18 \\ MPP-2018-35 }
\end{flushright}

\begin{center}
\vspace{1.5truecm}

{\Large \bf Generalised fluxes, Yang-Baxter deformations \vspace*{0.2cm} \\ and the O(d,d) structure of non-abelian $T$-duality}

\vspace{2truecm}

{Dieter L\"ust and David Osten}

\vspace{1.0truecm}

{\em Arnold Sommerfeld Center for Theoretical Physics,\\
LMU, Theresienstraße 37, 80333 M\"unchen, Germany}

\vspace{0.4truecm}

{\em Max-Planck-Institut f\"ur Physik \\
F\"ohringer Ring 6, 80805 M\"unchen, Germany}

\vspace{1.0truecm}

{{\tt dieter.luest@lmu.de, \quad ostend@mppmu.mpg.de}}

\vspace{1.5truecm}
\end{center}

\begin{abstract}
\begin{normalsize}
Based on the construction of Poisson-Lie $T$-dual $\sigma$-models from a common parent action we study a candidate for the non-abelian respectively Poisson-Lie $T$-duality group. This group generalises the well-known abelian $T$-duality group O$(d,d)$ and we explore some of its subgroups, namely factorised dualities, $B$- and $\beta$-shifts. The corresponding duality transformed $\sigma$-models are constructed and interpreted as generalised (non-geometric) flux backgrounds. 

We also comment on generalisations of results and techniques known from abelian $T$-duality. This includes the Lie algebra cohomology interpretation of the corresponding non-geometric flux backgrounds, remarks on a double field theory based on non-abelian $T$-duality and an application to the investigation of Yang-Baxter deformations. This will show that homogeneously Yang-Baxter deformed $\sigma$-models are exactly the non-abelian $T$-duality $\beta$-shifts when applied to principal chiral models. 
\end{normalsize}
\end{abstract}

\newpage
\tableofcontents

\newpage
\section{Introduction}
Dualities are a key aspect of quantum field and string theory, as they connect seemingly different theoretical settings and can be very useful tools to approach otherwise inaccessible problems. An important example for such dualities is target space or $T$-duality of two dimensional $\sigma$-models. In its simplest and most rigorous setting the duality maps two toroidal fibre bundles as target spaces to each other. This duality is based on the global abelian isometries of the target spaces, hence called abelian $T$-duality, and extends to the quantum theory (for reviews see \cite{Giveon1994a,Alvarez1995}). On the classical level, the reasoning behind this does not only work for $($U$(1))^d$-isometries of the target space, but also for a generic global group isometry, although the dual model does not possess this isometry anymore. This is known as non-abelian $T$-duality \cite{DeLaOssa1993}.

Klim\v{c}ik and \v{S}evera \cite{Klimcik1995,Klimcik1996} found a (non-isometric) generalisation of this setup, which also admits a version of target space duality named Poisson-Lie $T$-duality. The corresponding $\sigma$-models, called Poisson-Lie $\sigma$-models in the following, are the objects of study of this article and can be put into the form
\begin{large}
\begin{align*}
S  = \frac{1}{2} \int \mathrm{d}^2 \sigma \ (g^{-1} \partial_+ g)^a \left(\frac{1}{\frac{1}{G_0 + B_0} + \Pi(g)} \right)_{ab} (g^{-1} \partial_- g)^b
\end{align*}
\end{large}
with Lie group $\mathcal{G}$-valued fields $g$, a Poisson bivector $\Pi(g)$ on $\mathcal{G}$ and constant matrices $G_0$, $B_0$, describing metric and $B$-field evaluated at the identity of $\mathcal{G}$. The basic algebraic structure of this $\sigma$-model is a bialgebra, given as a so-called Manin triple, $\mathfrak{d} = \mathfrak{g} \oplus_{\mathfrak{d}} \mathfrak{g}^\star$, where $\mathfrak{g}$ is the $d$-dimensional Lie algebra of $\mathcal{G}$ and $\mathfrak{g}^\star$ is the dual algebra to $\mathfrak{g}$ determined by the Poisson structure $\Pi$. The bialgebra $\mathfrak{d}$ admits an O$(d,d)$-metric corresponding to the canonical pairing of $\mathfrak{g}$ and is defined by the fact, that this O$(d,d)$-metric is Ad-invariant. $\mathcal{G}$-isometric models and their non-abelian $T$-duals are obtained from the above action by choosing $\mathfrak{d}$ to be semi-abelian ($\mathfrak{g}$ or $\mathfrak{g}^\star$ is abelian). The toroidal models admitting standard abelian $T$-duality are of course also contained by choosing $\mathcal{G}$ to be abelian.

Non-abelian and Poisson-Lie $T$-duality does in general not extend to the quantum level and in some cases even spoils conformal invariance \cite{Giveon1994b,Elitzur1995,Tyurin1996,Klimcik1996c,Parkhomenko1999,Sfetsos1998,Sfetsos2009}. Nonetheless the study of the corresponding $\sigma$-model revealed geometric structures behind them and was also successfully used as a solution generating technique in supergravity \cite{Sfetsos2011a,Lozano2011,Itsios2012,Macpherson2013,Itsios2013a,Itsios2013,Lozano2014}.

Moreover (non-)abelian $T$-duality became of interest not only as generating technique in the context of supergravity solutions, but was also useful in the field of integrable models. Non-abelian $T$-duality and the algebraic structures behind Poisson-Lie $\sigma$-models inspired different kinds of integrable deformed $\sigma$-models, namely $\lambda$- and Yang-Baxter deformations \cite{Klimcik2008,Delduc2013b,Sfetsos2014}. The latter seem to be closely related to certain non-abelian $T$-duality transformations in general \cite{Kawaguchi2014a,Orlando2016a,Borsato2016,Borsato2016b,Hoare2016,Hoare2016c}. In particular for so-called abelian $r$-matrices one reproduces $\beta$-shifts of (abelian) $T$-duality \cite{Osten2017}. A clarification of the appropriate generalisation of this statement to the non-abelian case was one of the motivations of this article and we will comment on this aspect extensively. But also in general, the study of dualities of two dimensional integrable $\sigma$-models and their geometric and algebraic description should be useful to reveal the symmetries behind integrability there.

This paper aims to introduce techniques of abelian $T$-duality also in context of its non-abelian generalisation. This includes the investigation of generalised fluxes of the connected backgrounds, the above mentioned application to integrable deformations and the systematic study of a duality group, which motivates to look for a double field theory making such a duality group manifest.

\pagebreak

\subsection*{Results and overview}
After setting up our conventions and reviewing basics of $T$-duality and bialgebras, we revisit the definition \cite{Klimcik1995} of the non-abelian $T$-duality (NATD) group to a Poisson-Lie $\sigma$-model associated to the bialgebra $\mathfrak{d} = \mathfrak{g} \oplus_{\mathfrak{d}} \mathfrak{g}^\star$
\begin{align*}
\text{NATD group}(\mathfrak{d}) = \left\lbrace \text{Manin triple decompositions of } \mathfrak{d} \right\rbrace 
\end{align*}
and motivate it from the construction of the Poisson-Lie $\sigma$-model from a doubled $\sigma$-model. The central observation of this article is, that the elements of this group are \textit{vector space} automorphisms $\varphi$ of $\mathfrak{d}$, preserving the O$(d,d)$-metric and the algebraic closure of $\mathfrak{g}$ and $\mathfrak{g}^\star$,
\begin{align*}
[\varphi(\mathfrak{g}),\varphi(\mathfrak{g})]\subset \varphi(\mathfrak{g}), \qquad [\varphi(\mathfrak{g}^\star),\varphi(\mathfrak{g}^\star)]\subset \varphi(\mathfrak{g}^\star)
\end{align*} 
and that we can thus analyse this group with help of the usual decomposition of an O$(d,d)$-transformation into factorised dualities, $B$-shifts, $\beta$-shifts and $GL$-transformations fulfilling these additional closure conditions. This gives insights into (at least parts of) the duality group. The analysis shows that the above definition seems to be too narrow in some cases - this is demonstrated by constructing the non-abelian $T$-dual of a principal chiral model w.r.t. a subgroup. A less restrictive version of the Poisson-Lie $\sigma$-model and the NATD group is proposed, where closure of $\mathfrak{g}$ resp. $\varphi(\mathfrak{g})$ is not required anymore, such that the above example it contained. For $B$- and $\beta$-shifts of O$(d,d)$ to lie in the NATD group they have to fulfil non-trivial conditions, depending on the algebraic structure of $\mathfrak{d}$. In some simple cases we can directly construct the corresponding transformations from the $\sigma$-model, although generically this is not possible. One of these cases is a setting which admits a general version of a generalised Buscher procedure for $\beta$-transformations, already introduced in \cite{Hoare2016,Hoare2016c}.

After reviewing generalised fluxes we demonstrate that Poisson-Lie $\sigma$-models are non-trivial realisations of a string in a generalised flux background. Subsequently we are able to interpret the Bianchi identities of these generalised fluxes in our special case - ${\mathbf{f}^c}_{ab}$ are Lie algebra structure constants - in terms of Lie algebra cohomology of $\mathfrak{g}$ and show that the non-geometric $\mathbf{Q}$-flux can be interpreted as structure constants of a Lie algebra dual to $\mathfrak{g}$, with a its non-associativity (violation of its Jacobi identity) generated by an $\mathbf{R}$-flux. In terms of Lie algebra cohomology on $\mathfrak{g}$ the non-geometric part of Poisson-Lie $\sigma$ is described by 
\begin{align*}
\mathbf{Q} \in H^1(\mathfrak{g},\mathfrak{g}\wedge\mathfrak{g}), \quad \delta\mathbf{R} \in [0] \in H^1(\mathfrak{g},\mathfrak{g}\wedge\mathfrak{g}\wedge\mathfrak{g}),
\end{align*}
generalising the abelian case, studied in \cite{Bakas2013}.

The previous considerations resemble ones in double field theory and generalised geometry. In particular the construction of the Poisson-Lie $\sigma$-model from a doubled $\sigma$-model has the geometric interpretation of a projection onto Dirac subbundles of the bialgebra respectively its group, the so-called Drinfel'd double. We comment on different splitting structures defining such Dirac decompositions, analyse a few candidates for these - almost para-complex and symplectic structures - and comment on their integrability, which is analogue to the algebraic closure conditions on the non-abelian $T$-duality group.

Finally we include an interpretation of Yang-Baxter deformed backgrounds as non-geometric flux backgrounds and also a proof of equivalence between homogeneous Yang-Baxter deformations and $\beta$-shifts of this non-abelian $T$-duality group.

For the convenience of the reader we include an appendix with the basic notions of (Chevalley-Eilenburg) Lie algebra cohomology and an overview between the connection of solutions to Yang-Baxter equations and bialgebras.

\pagebreak
\section{Poisson-Lie $T$-duality and $\sigma$-models}
\label{chap:Review}
\subsection{Motivation from the toroidal case}
In this section we first introduce some aspects of (classical) abelian $T$-duality, such as the Buscher procedure and the duality group O$(d,d)$, in order to motivate their equivalents in the non-abelian case. We then review a particularly interesting formulation on $\beta$-shifts via non-abelian $T$-duality, that will be of use later in this paper.

\subsubsection{Abelian $T$-duality}
The non-linear $\sigma$-model of world-sheet embeddings into an $n$-dimensional target space is\footnote{Here and in the following we will neglect the dilaton and its behaviour under $T$-duality because we are mainly interested in the classical case. The dilaton couples at higher order in $\alpha^\prime$ to the worldsheet and its change under $T$-duality comes from a transformation of the path integral. 

Throughout this paper we use the following conventions: The world-sheet coordinates are $\tau$ and $\sigma$ with world-sheet metric in conformal gauge $\gamma_{\alpha \beta} = \text{diag}(-1,1)$. We also use $\dot{Y} = \partial_\tau Y$ and $Y^\prime = \partial_\sigma Y$ and the light-cone coordinates $\sigma^\pm = \tau \pm \sigma$, such that $\star \mathrm{d}\sigma^{\pm} = \pm  \mathrm{d}\sigma^{\pm}$.} 
\begin{equation}
S = \frac{1}{2} \int \left[ G_{ij}(X) \mathrm{d} X^i \wedge \star \mathrm{d} X^j + B_{ij}(X) \mathrm{d} X^i \wedge \mathrm{d} X^j \right], \qquad i,j=1,...,n \label{eq:ActionNLSM}
\end{equation}
with a target space metric $G(X)$ and Kalb-Ramond field $B(X)$. Now assume the target space has an isometry and choose coordinates $X^i = (X^1,X^{\underline{i}})$ such that $X^1$ parameterises the isometry. Using this fact we can rewrite the action by substituting $\mathrm{d}X^1$ by gauge fields $A$. For this we also have to add a Lagrangian multiplier term $-\bar{X}_1 \mathrm{d}A$ to the action, such that the new action is classically equivalent to \eqref{eq:ActionNLSM} by enforcing that $A$ is flat, $A = \mathrm{d}X^1$. Integrating out the gauge field $A$ instead of the Lagrangian multiplier $\bar{X}_1$ we get the new $\sigma$-model of the form
\begin{equation}
S = \frac{1}{2} \int \left[ \bar{G}^{ij}(\bar{X}_{\underline{i}}) \mathrm{d} \bar{X}_i \wedge \star \mathrm{d} \bar{X}_j + \bar{B}^{ij}(\bar{X}_{\underline{i}}) \mathrm{d} \bar{X}_i \wedge \mathrm{d} \bar{X}_j \right]
\end{equation}
with coordinates $\bar{X}_i = (\bar{X}_1,X^{\underline{i}})$ where the new background $\bar{E} = \bar{G} + \bar{B}$ for the metric $\bar{G}$ and the Kalb-Ramond field $\bar{B}$ is given by the so-called Buscher rules
\begin{align}
\bar{G}^{11} &= \frac{1}{G_{11}}, \quad \bar{E}^{1 \underline{m}} = \frac{E_{1 \underline{m}}}{E_{11}}, \quad \bar{E}^{\underline{m} 1} = - \frac{E_{\underline{m}1}}{E_{11}}, \quad \bar{E}^{\underline{m} \underline{n}} = E_{\underline{m} \underline{n}}  - \frac{E_{\underline{m}1} E_{1 \underline{n}} }{E_{11}} \label{eq:BuscherRules}
\end{align}
in terms of the old background $E = G+B$. In case there are $d$ such U$(1)$-isometries, or in other words the target space is a toroidal fibration, we can choose coordinates such these fibres are parameterised by the $X^i = (X^\alpha , X^{\underline{\alpha}})$ and apply the above procedure. Combining the above discussed $\mathbb{Z}_2$-dualities with GL$(d)$-transformations of the $X^\alpha$ yields the duality orbit
\begin{equation}
\bar{E} = \varphi.E = (aE + b)(cE+d)^{-1}. \label{eq:OddAction}
\end{equation}
where $a = \text{diag}(A,\mathbb{1})$, $b= \text{diag}(b,0)$, $c= \text{diag}(c,0)$ and $d = \text{diag}(D,\mathbb{1})$ are each $n$ by $n$ matrices and
\begin{align*}
\varphi = \left( \begin{array}{cc} A& B \\ C&D \end{array} \right) \quad \in \text{O}(d,d),
\end{align*}
where O$(d,d)$ is defined w.r.t. the metric $\eta = \left( \begin{array}{cc} & \mathbb{1} \\ \mathbb{1} &
\end{array} \right)$.
The duality group is O$(d,d)$ and can be revealed only by considering the quadratic part in derivatives of the isometry coordinates - the transformation of non-isometry coordinates (also called spectators sometimes) is easily reproduced from \eqref{eq:OddAction}, thus these are normally neglected and we will follow this route for most parts of this paper. Later in this article we will see that this separation needs to be treated with a little care in the non-abelian generalisation.

Let us briefly review the structure of this duality group. Any $\varphi \in$O$(d,d)$ can be generated by elements of the following four subgroups.
\begin{itemize}
\item The true $\mathbb{Z}_2$-dualities, corresponding to the Buscher rules, are normally called \textit{factorised dualities}. They fulfil $\varphi^2 = \mathbb{1}$ and 'generate' the components of O$(d,d)$, that are not connected to the identity:
\begin{equation}
\varphi_{T_i} = \left( \begin{array}{cc} \mathbb{1} - E_{ii} & E_{ii}  \\ E_{ii} & \mathbb{1} - E_{ii} \end{array} \right) \qquad \text{with} \qquad (E_{ij})_{kl} = \delta_{ik}\delta_{jl}.
\end{equation}

\item \textit{General linear transformations} $G + B \rightarrow A^T (G+B) A$ are contained in this representation of O$(d,d)$ as
\begin{equation}
\varphi_{GL} = \left( \begin{array}{cc} A^T&  \\ & A^{-1} \end{array} \right) \qquad \text{for} \qquad A \in \text{GL}(d).
\end{equation}

\item \textit{$B$-shifts} by a constant skewsymmetric matrix also form a subgroup of the duality group. They correspond to gauge transformations of the $\mathbf{H}$-flux, $\mathbf{H}=\mathrm{d}B$. 

\item Performing a 'full' factorised duality $\varphi = \eta$, gives a new background $\bar{E} = g + \beta$, where $\beta$ takes the role of $B$ and is given by
\begin{equation}
\beta = - (G+B)^{-1} B (G-B)^{-1}.
\end{equation} 
We will refer to $\beta$ as being dual or conjugate to $B$. So, logically \textit{$\beta$-shifts} by a skewsymmetric matrix form another subgroup of O$(d,d)$. We discuss their meaning on the $\sigma$-model level in the next paragraph.

The matrix representations of $B$- and $\beta$-shifts are given by
\begin{equation}
\varphi_{B} = \left( \begin{array}{cc} \mathbb{1} & b  \\ & \mathbb{1} \end{array} \right), \qquad \varphi_{\beta} = \left( \begin{array}{cc} \mathbb{1} &  \\ r & \mathbb{1} \end{array} \right)
\end{equation}
for skewsymmetric $b$ and $r$.
\end{itemize}
The O$(d,d)$ duality group cannot only be motivated as above but also from the Hamiltonian. With the canonical momentum
\begin{equation}
P_i = G_{ij} \dot{X}^j + B_{ij} X^{\prime j}
\end{equation}
we can compute a first order form of the action:
\begin{align}
\mathcal{L} &= \dot{X}^i P_i - H \label{eq:AbelianDoubledLag} \\
\text{with} \quad H &= ( X^\prime , P ) \left( \begin{array}{cc} G - BG^{-1} B & BG^{-1} \\ - G^{-1} B & G^{-1} \end{array} \right) \left( \begin{array}{c} X^\prime \\ P \end{array} \right). \label{eq:AbelianHamiltonianDensity}
\end{align}
The Hamiltonian density $H$ is invariant under linear transformations by $\varphi \in$O$(d,d)$ of $(X^\prime,P)$ and of the so-called generalised metric
\begin{equation}
\mathcal{H}(G,B) =  \left( \begin{array}{cc} G - BG^{-1} B & BG^{-1} \\ - G^{-1} B & G^{-1} \end{array} \right). \label{eq:GeneralisedMetric}
\end{equation}
From a similar first order form of an action we will motivate the non-abelian generalisation of the duality group O$(d,d)$.

More details on abelian $T$-duality, including comments on the issues on integrating the gauge fields out and the discussion to the quantum case, where the duality group reduces to its discrete subgroup O$(d,d;\mathbb{Z})$, can be found for example in \cite{Rocek1992,Buscher1988,Tseytlin1990,Tseytlin1991,Giveon1992,Siegel1993,Giveon1994,Alvarez1994,Alvarez1994a,Alvarez1995,Alvarez1996,Alvarez2000,Bouwknegt2004,Plauschinn2014,Rennecke2014}.

\subsubsection{$\beta$-shifts via a generalised Buscher procedure}
\label{chap:betaTransformationAbelian}
The $\sigma$-model interpretation of $\beta$-shifts will be of special interest in the following. We will show two points of view on $\beta$-shifts. Starting from the linear $\sigma$-model
\begin{equation}
S = \frac{1}{2} \int \left[ G_{ij} \mathrm{d} X^i \wedge \star \mathrm{d} X^j + B_{ij} \mathrm{d} X^i \wedge \mathrm{d} X^j \right] \label{eq:ActionTorus}
\end{equation}
for \textit{constant} metric $G$ and Kalb-Ramond field $B$, we follow the steps
\begin{enumerate}
\item gauging all the U$(1)$-isometries: substituting $\mathrm{d} X^i$ by gauge fields $A^i$ and introducing a Lagrangian multiplier term $\bar{X}_i \wedge A^i$ in the Lagrangian,
\item $B$-shift in the dual picture: adding a term $\bar{X}^* \beta = \beta^{ij} \mathrm{d}\bar{X}_i \wedge \mathrm{d} \bar{X}_j$ with constant and skewsymmetric coefficients $\beta^{ij}$, which is a total derivative as $\mathrm{d} \beta = 0$, to the Lagrangian,
\item step 1 for the U$(1)$-isometries of the dual coordinates $\bar{X}_i$,
\end{enumerate}
we arrive at
\begin{equation}
S^\prime = \frac{1}{2} \int A^i \wedge \left( G_{ij} \star A^j + B_{ij} A^j \right) + \bar{A}_i \wedge \left( \beta^{ij}\bar{A}_j + A^i \right) + \mathrm{d} X^i \wedge \bar{A}_i. \label{eq:ActionGaugedBetaShifted}
\end{equation}
Integrating out $A^i$ and $\bar{A}_i$ gives the standard O$(d,d)$ $\beta$-shift\footnote{In the literature to integrable deformations $\beta$-shifts are usually referred to by $TsT$-transformation, standing for $T$-duality - coordinate \textit{s}hift - $T$-duality. See for example the original papers \cite{Frolov2005,Alday2006} and proof of the equivalence between $\beta$-shifts and abelian Yang-Baxter deformations \cite{Osten2017}.}
\begin{align}
S &= \frac{1}{2} \int \mathrm{d} X^i \wedge \left[ \tilde{G}_{ij} \star  + \tilde{B}_{ij} \right] \mathrm{d} X^j = \int \mathrm{d}^2 \sigma \ \partial_+ X^i \left(\frac{1}{\frac{1}{G+B} + \beta}\right)_{ij} \partial_- X^j \label{eq:ActionTorusBetaShifted}.
\end{align}
The second perspective on a $\beta$-shift, that we will take, is the following. A $\beta$-shift cannot only be interpreted as a $B$-shift in the dual coordinates, but also as the introduction of a Poisson bivector $\Pi = \beta^{ij} \partial_i \wedge \partial_j$. On a symplectic leaf\footnote{so choosing and restricting to coordinates such that $\left(\beta^{-1}\right)_{ij}$ exists. We define $\left(\beta^{-1}\right)_{ij}$ to vanish on all other coordinates.} of $\Pi$, we can define a two-form \linebreak
$\omega = \beta^{-1}_{ij} \mathrm{d} X^i \wedge \mathrm{d} X^j$, which is symplectic for constant $\beta^{ij}$:
\begin{equation}
\left[\Pi , \Pi\right]_S = 0 \quad \Leftrightarrow \quad \mathrm{d}\omega = 0, \label{eq:CocycleAbelian}
\end{equation}
where $[ \ , \ ]_S$ is the Schouten bracket of multivectors. On such a symplectic leaf and after integrating out the $\bar{A}_i$ in \eqref{eq:ActionGaugedBetaShifted} first and redefining $A^j \mapsto A^j -\mathrm{d} X^j$, we get
\begin{equation}
S = \frac{1}{2} \int \mathrm{\textbf{D}} X^i \wedge \left( G_{ij} \star \mathrm{\textbf{D}} X^j + B_{ij} \mathrm{\textbf{D}} X^j \right) + \beta^{-1}_{ij} A^i \wedge A^j \label{eq:ActionGaugedBetaShifted2}
\end{equation}
with $\mathrm{\textbf{D}}X^i = \mathrm{d} X^i - A^i$. With the identification 
\begin{equation}
\mathrm{d}\bar{X}_i = G_{ij} \star \mathrm{d} X^j + B_{ij} \mathrm{d} X^j
\end{equation}
between dual (winding) and the original coordinates and subtraction of a total derivative $\mathrm{d} \bar{X}_i \wedge \mathrm{d} X^i$ we obtain
\begin{equation}
S = \frac{1}{2} \int A^i \wedge \left( G_{ij} \star A^j + B_{ij} A^j \right) + \mathrm{d} \bar{X}_i \wedge A^i + \beta^{-1}_{ij} A^i \wedge A^j . \label{eq:ActionGaugedBetaShifted3}
\end{equation}

The reason, that we consider the version \eqref{eq:ActionGaugedBetaShifted3} of \eqref{eq:ActionGaugedBetaShifted}, is that it can be obtained via a different route, which was introduced in \cite{Hoare2016}:
\begin{itemize}
\item Given a cocycle $\omega$ \eqref{eq:CocycleAbelian} we can centrally extend the isometry algebra with generators $\{ t_i \}$ (abelian in the toroidal case) in the following way (see appendix \ref{chap:Cohomology} for more details), where $Z$ is the new central element of the algebra:
\begin{align}
[t_i,t_j] =  0 \quad \mapsto \quad [t_i , t_j]^\prime &= [t_i , t_j] + \omega(t_i , t_j) Z = \beta^{-1}_{ij} Z, \label{eq:CentralExtensionAbelian} \quad [ t_i , Z ]^\prime = 0. 
\end{align}

\item Starting from original action \eqref{eq:ActionTorus}, we substitute $\mathrm{d} X^i$ by gauge fields $A^i$, which we now assume to be components of a gauge field $A^\prime = A^i t_i +C Z$ with field strength $F^\prime = \mathrm{d} A^\prime - [A^\prime\overset{\wedge}{,}A^\prime]^\prime$. Using \eqref{eq:CentralExtensionAbelian} the components of $F$ are
\begin{equation}
F^i = \mathrm{d}A^i \quad \text{and} \quad F^Z = \mathrm{d}C - \beta^{-1}_{ij} A^i \wedge A^j.
\end{equation}
Instead of adding a Lagrangian multiplier term which enforces only $F^i = \mathrm{d} A^i = 0$, we also want to set $F^Z = 0$. For this we use 'extended' dual coordinates $Y_s = \left(\bar{X}_i,Y\right)$ and add the following term to the Lagrangian
\begin{align}
\mathcal{L}_{Lag. mult.}  &\propto - Y_s F^s \overset{P.I.}{=} \mathrm{d} \bar{X}_i \wedge A^i + \mathrm{d} Y \wedge C + Y \beta^{-1}_{ij} A^i \wedge A^j.
\end{align}
\item Integrating out $C$ leads to $Y = const.$, so that the resulting action is the same as \eqref{eq:ActionGaugedBetaShifted3}. After integrating out $A$ we are left with \eqref{eq:ActionTorusBetaShifted}.
\end{itemize}
All these manipulations were rather trivial in the abelian case, but this analysis helps to understand the geometrical meaning and the objects to look for in the non-abelian case. We will comment on this and the connection to Lie algebra cohomology in section \ref{chap:NATDsubgroups}.

\subsubsection{Non-abelian $T$-duality}
Here we show that also based on non-abelian isometries a similar gauging procedure can be done. In general this is known as non-abelian $T$-duality (NATD in the following). Let us consider the model
\begin{equation}
S = \frac{1}{2} \int \mathrm{d}^2 \sigma \ (g^{-1} \partial_+ g)^a E_{ab} (g^{-1} \partial_- g)^b \label{eq:NATDoriginal}
\end{equation}
with constant 'background' field $E = G+B$ and group valued fields $g$: $\Sigma \rightarrow \mathcal{G}$, with corresponding Lie algebra $\mathfrak{g}$ and structure constants ${f^c}_{ab}$. Here we neglect as discussed above additional spectator coordinates. We perform a procedure, which is very similar to the abelian case: Substitute $g^{-1} \mathrm{d} g$ by $\mathfrak{g}$-valued gauge fields $j$, add a Lagrangian multiplier term \linebreak $x_a (\mathrm{d}j + [j,j])^a$, which fixes $j$ to be pure gauge and then integrate out $j$. The non-abelian $T$-dual model is
\begin{equation}
S = \frac{1}{2} \int \mathrm{d}^2 \sigma \ \partial_+ x_a \bar{E}^{ab}(x) \partial_- x_b, \qquad \bar{E}^{-1}_{ab} = E_{ab} - x_c {f^c}_{ab}. \label{eq:NATDdual}
\end{equation}
So in contrast to the abelian case, the duality connects an isometric and a non-isometric model with each other. In general the status of non-abelian $T$-duality is not as strong as the one of abelian $T$-duality - it is not supposed to be a true duality on the quantum level, but a map between similar theories. Also if the trace of the structure constants of $\mathfrak{g}$ does not vanish, ${f_c}^{ac} \neq 0$, the non-abelian $T$-dual model \eqref{eq:NATDdual} possesses a kind of anomaly that spoils conformal invariance in the quantum theory. For more details on non-abelian $T$-duality see \cite{DeLaOssa1993,Gasperini1993,Alvarez1994b,Giveon1994b,Elitzur1995,Hoare2016,Alvarez1994}.
 
\subsection{Lie bialgebras}
\label{chap:BialgebraDef}
Before we discuss the further generalisation of the above, i. e. Poisson-Lie $T$-duality and Poisson-Lie $\sigma$-models, which will also enlighten the structure of the action \eqref{eq:NATDdual}, let us review the algebraic basics for this - Lie bialgebras - and set up our conventions.

Here and in the rest of the paper, we consider a $d$-dimensional semi-simple\footnote{In principle we do not have to restrict to the semi-abelian case to perform non-abelian $T$-duality. This was successfully demonstrated in \cite{Hong2018}.} Lie algebra $\mathfrak{g}$ with corresponding Lie group $\mathcal{G}$, the Killing form $\kappa$ and generators $t_a$ fulfilling
\begin{equation}
[ t_a , t_b ] = {f^c}_{ab} t_c .
\end{equation}
We use $\partial_a,\partial_b,...$ to represent the curved derivatives corresponding to $t_a,t_b,...$ treated as invariant vector fields on $\mathcal{G}$, and $\partial_i,\partial_j,...$ for flat derivatives.

\subsubsection{Lie bialgebra definitions}
\textbf{Definition via Manin triples.} 
We want to define a Lie algebra bracket on the vector space $\mathfrak{g} \oplus \mathfrak{g}^*$ in terms  of the $2d$ generators $T_A = (t_a , \bar{t}^a)$ of $\mathfrak{g} \oplus \mathfrak{g}^*$, such that the canonical, non-degenerate and symmetric bilinear form, defined by
\begin{equation}
\braket{t_a | t_b} = \braket{\bar{t}^a |\bar{t}^b} = 0, \quad \braket{t_a | \bar{t}^b} = \delta_a^b \label{eq:BialgebraMetric}
\end{equation}
or in terms of the $T_A$, $\braket{T_A | T_B} = \eta_{AB}$ with $\eta = \left( \begin{array}{cc} & \mathbb{1}_d \\ \mathbb{1}_d & \end{array} \right)$, is Ad-invariant, i.e.
\begin{equation}
\braket{T_A | [T_B , T_C]} = \braket{[T_C,T_A] | T_B}.
\end{equation}
The structure constants of a complementary pair $(\mathfrak{g},\mathfrak{g}^\star)$ of Lagrangian (meaning maximally isotropic w.r.t. to $\langle \ \vert \ \rangle$) subalgebras can be constructed to be of the form
\begin{align}
[T_A , T_B] &= {F^C}_{AB} T_C \nonumber \\
\text{with} \quad [ t_a , t_b ]  &= {f^c}_{ab} t_c, \quad [ \bar{t}^a , \bar{t}^b] = {\overline{f}_c}^{ab} \bar{t}^c, \label{eq:Bialgebra}  \\
[ t_a , \bar{t}^b ]  &= {\overline{f}_a}^{bc} t_c + {f^b}_{ca} \bar{t}^c. \nonumber
\end{align}
The Lie group to the Lie algebra $\mathfrak{d}$ is called Drinfel'd double, we denote it by $\mathcal{D}$. It contains $\mathcal{G}$ and $\bar{\mathcal{G}}$ (the Lie group to $\mathfrak{g}^\star$) as subgroups, $\mathcal{D} = G \Join \bar{G}$. The condition on the structure constants ${f^c}_{ab}$ and ${\overline{f}_c}^{ab}$ in order for \eqref{eq:Bialgebra} to fulfil the Jacobi identity is\footnote{We use the following notations for antisymmetrisation of indices
\begin{equation}
u_{(a} v_{b)} = u_a v_b - u_b v_a \qquad \text{and} \qquad u_{(a|} v_{c|b)} = u_{a} v_{cb} - u_{b} v_{ca}. \nonumber
\end{equation}
}
\begin{equation}
{f^c}_{mn} {\overline{f}_c}^{ab}  ={f^{(a}}_{c(m}  {\overline{f}_{n)} } ^{b)c}. \label{eq:BialgebraJacobi}
\end{equation}
The triple $(\mathfrak{d},\mathfrak{g},\mathfrak{g}^\star)$ is called Manin triple. For a given $(\mathfrak{d} , \langle \ | \ \rangle)$ there can be multiple Manin triple decompositions. In the following we also use the notation
\begin{equation}
\mathfrak{g}\oplus_{\mathfrak{d}} \mathfrak{g}^\star
\end{equation}
to describe a Manin triple. Consistency requires that $\kappa^{-1}$ is the Killing form on $\mathfrak{g}^\star$.
\pagebreak \vspace*{0.01cm} \\
\textbf{Definition via 1-cocycles.} In the maths literature bialgebras are normally defined differently but of course equivalently. A bialgebra is the pair $(\mathfrak{g},u)$ of a Lie algebra $\mathfrak{g}$ and a $\mathfrak{g} \wedge \mathfrak{g}$-valued 1-cochain $u$ on $\mathfrak{g}$ fulfilling
\begin{enumerate}
\item 1-cocycle condition:
\begin{equation}
\delta u (m,n) := \Delta(\text{ad}_m) u(n) - \Delta(\text{ad}_n) u(m) - u([m,n]) = 0 \label{eq:Bialgebra1cocycle1}
\end{equation}
\item 'Jacobi identity':
\begin{equation}
\Delta(u) \circ u = 0, \label{eq:Bialgebra1cocycle2}
\end{equation}
\end{enumerate}
where we defined the coproduct $\Delta(X) := \mathbb{1} \otimes X + X \otimes \mathbb{1}$. A $\mathfrak{g}\wedge\mathfrak{g}$-valued 1-cochain
\begin{equation}
u(t_c) = {u_c}^{ab} t_a \wedge t_b
\end{equation}
can be identified with a skew-symmetric bracket on $\mathfrak{g}^\star$, $[ \ , \ ]_{\mathfrak{g}^\star} : \ \mathfrak{g}^\star \wedge \mathfrak{g}^\star \rightarrow \mathfrak{g}^\star$ with structure constants ${{\overline{f}}_c}^{ab} \equiv {u_c}^{ab}$. Then the 1-cocycle condition \eqref{eq:Bialgebra1cocycle1} is equivalent to \eqref{eq:BialgebraJacobi} and \eqref{eq:Bialgebra1cocycle2} corresponds to the Jacobi identity on $\mathfrak{g}^\star$. So indeed, this definition is equivalent to the Manin triple.

If the 1-cocycle is a 1-coboundary, we will call the corresponding bialgebra \textit{'1-coboundary'}. We will comment on a certain type of these 1-coboundary bialgebras in the next section. But, of course, there are more possible bialgebras. In appendix \ref{chap:WithoutYB} we comment on this.

\subsubsection{$R$-brackets and Yang-Baxter equations}
Given a Lie algebra $\mathfrak{g}$ with bracket $[ \ , \ ]$, is it possible to define another Lie bracket on $\mathfrak{g}$? A simple candidate is the so-called $R$-bracket
\begin{align}
[m,n]_R = [m,R(n)] - [n,R(m)] \qquad \forall m,n \in \mathfrak{g} \label{eq:Rbracket}
\end{align}
for some $R \in \text{End}(\mathfrak{g})$. A sufficient condition on $R$, s.t. $[ \ , \ ]_R$ fulfils the Jacobi-identity, is
\begin{equation}
[R(m),R(n)] - R([m,n]_R) = c^2 [m,n], \qquad \forall m,n \in \mathfrak{g}. \label{eq:mCYBE}
\end{equation}
This condition can be rewritten as
\begin{equation}
[(R\pm c \mathbb{1})(m),(R\pm c\mathbb{1})(n)] = (R\pm c\mathbb{1})([m,n]_R), \qquad \forall m,n \in \mathfrak{g}, \label{eq:mCYBEhomomorphism}
\end{equation}
which means that $(R \pm c\mathbb{1})$ is a Lie algebra homomorphism between $(\mathfrak{g},[ \ , \ ])$ and \\ $\mathfrak{g}_R := (\mathfrak{g},[ \ , \ ]_R)$.
After rescaling we can distinguish three cases of \eqref{eq:mCYBE}:
\begin{enumerate}
\item $c = 0$: classical Yang-Baxter equation (cYBe),
\item $c = 1$: \textit{non-split} modified classical Yang-Baxter equation (mcYBe)
\item $c = i$: \textit{split} modified classical Yang-Baxter equation,
\end{enumerate}
each of which have distinct roles in the definition of Lie bialgebra structures of semisimple Lie algebras, which is sketched in appendix \ref{chap:Rbrackets}. A more extensive review can be found for example in \cite{Vicedo2015}.

The connection of $R$-brackets of semisimple Lie algebras to their bialgebra structures is seen via the definition via 1-cocycles. 
The non-degenerate Killing form $\kappa$ on $\mathfrak{g}$ defines a 2-vector $r = r^{ab} t_a \wedge t_b$ for each $R \in \text{End}(\mathfrak{g})$, $r^{ab} = \kappa^{ac} {R^b}_c$. The $\mathfrak{g} \wedge \mathfrak{g}$-valued 1-coboundary
\begin{align}
\delta r (x) = \Delta(\text{ad}_x) r = [\Delta(x),r]
\end{align}
is trivially a 1-cocycle \eqref{eq:Bialgebra1cocycle1} and the condition for the Jacobi identity \eqref{eq:Bialgebra1cocycle2} on $r$ can be written as
\begin{equation}
\Delta(\text{ad}_x)\left([r,r]_S \right) = 0,
\end{equation}
where $[ \ , \ ]_S$ is the standard Schouten bracket of multivectors.

\subsubsection{Bivector fields on Lie groups}
In the following we review typical bivectors on Lie groups associated to a dual Lie algebra structure by the canonical relation
\begin{equation}
{\overline{f}_c}^{ab} = \partial_c \Pi^{ab}(e).
\end{equation}

\subsubsection*{Poisson bivectors}
The canonical Poisson vector $\Pi = \Pi^{ab} t_a \wedge t_b$ on a Lie group $\mathcal{G}$ for a left-(right-)invariant basis $\{t_a\}$ of $T \mathcal{G}$ is given by
\begin{equation}
\Pi^{ab}\left(g\right) = {\overline{f}_c}^{ab} x^c - \frac{1}{2} {\overline{f}_c}^{k(a} {f^{b)}}_{dk} x^c x^d + ... \ , \quad \text{for} \ g =\exp(x^a t_a) \in \mathcal{G}, \label{eq:PoissonBivectorHomogeneous}
\end{equation}
if the structure constants $f$, $\bar{f}$ describe a Lie bialgebra.\footnote{E.g. $\mathcal{O}(x)$ of $\Pi^{k(a} \partial_k \Pi^{bc)}$ vanishes because of the Jacobi identity on $\mathfrak{g}^\star$, all the higher order terms because of Jacobi identity on $\mathfrak{g}^\star$ and 1-cocycle condition \eqref{eq:1CocycleCondition}.} The explicit form of \eqref{eq:PoissonBivectorHomogeneous} can be derived from (formally) transporting ${\overline{f}_c}^{ab}x^c$ along $\mathcal{G}$. Let us note, that $\Pi$ in \eqref{eq:PoissonBivectorHomogeneous} is neither left- nor right-invariant, which allows for the fact, that $\Pi(e)=0$. For this reason we will call $\Pi$ \textit{homogeneous} Poisson structure in the following. The Poisson bivector \eqref{eq:PoissonBivectorHomogeneous} can be constructed also in a coordinate independent manner. Given a Lie bialgebra $\mathfrak{d}=\mathfrak{g} \oplus_{\mathfrak{d}} \mathfrak{g}^\star$ and the adjoint action of a $g \in \mathcal{G}$ on the generators of $\mathfrak{d}$ we can write the homogeneous Poisson bivector as \cite{Klimcik1995}
\begin{align}
\Pi(g) &= C(g) \cdot A^{-1} (g) \label{eq:PoissonHomogenousfromAd} \\
\text{for} \quad \text{Ad}_g T_A &= \left(\begin{array}{c} gt_a g^{-1} \\ g\bar{t}^a g^{-1} \end{array} \right) = {\left( \begin{array}{cc} A(g) & 0 \\ C(g) & (A^T)^{-1}(g) \end{array} \right)^B}_A T_B. \nonumber
\end{align}
From the properties of $\text{Ad}_g$ on $\mathfrak{d}$ we can deduce the useful relation
\begin{equation}
\partial_a \Pi^{bc}(g) = {\barf_a}^{bc} + {f^{(b}}_{ad} \Pi^{c)d}(g) \label{eq:PoissonHomogeneousDerivative}
\end{equation}
for $g=\exp(x^a t_a)$.

\subsubsection*{Invariant bivectors}
In case we have a 1-coboundary bialgebra associated to $\mathfrak{g}$ and the 0-cocycle $r = r^{ab} t_a \wedge t_b$ on $\mathfrak{g}$, we can define a non-homogeneous bivector
\begin{equation}
\Pi_r^{ab}\left(g\right) = r^{ab} - {\overline{f}_c}^{ab} x^c + ... \ , \quad \text{for} \ g =\exp(x^a t_a) \in \mathcal{G}, \label{eq:BivectorCYBE}
\end{equation}
which is simply $r$ transported via left-(right) translation along $\mathcal{G}$ and the same as \eqref{eq:PoissonBivectorHomogeneous} plus a constant term.
%This bivector field is left(right)-invariant given in a left(right)-invariant basis.\footnote{$\Pi^{ab} = r^{ab}$ is a right- resp. left-invariant version of the bivector in a left- resp. right-invariant basis.}

Generically $\Pi_r$ given by \eqref{eq:BivectorCYBE} will not be a Poisson structure, but the bialgebra properties on the structure constants $f$ and $\bar{f}$ mean that the Jacobiator will be constant and Ad-invariant. So for example, if $r$ corresponds to a solution of the (modified) classical Yang-Baxter equation, then
\begin{equation}
\Pi_r^{k(a} \partial_k \Pi_r^{bc)}(g)= c^2 \kappa^{am} \kappa^{bn} {f^c}_{mn}, 
\end{equation}
where $c^2$ is the coefficient in the (modified) classical Yang-Baxter equation \eqref{eq:mCYBE}. In case, $r$ is a solution of the classical Yang-Baxter equation ($c^2 = 0$), \eqref{eq:BivectorCYBE} is the left(right)-invariant non-homogeneous Poisson bivector, introduced by \cite{Drinfeld1983}.

For such 1-coboundary bialgebras we can reproduce the homogeneous Poisson bivector \eqref{eq:PoissonBivectorHomogeneous} from
\begin{equation}
\Pi = r - \Pi_{r}.
\end{equation}
This has been noted already in \cite{Hoare2017}. But as commented above, $\Pi^{ab}_r(e) = r^{ab}$ does not necessarily have to be a solution to a Yang-Baxter equation in order for $\Pi$ to be Poisson.

\subsection{Construction of Poisson-Lie $\sigma$-models from a doubled $\sigma$-model}
Let us review the basic construction und geometry of Poisson-Lie $\sigma$-models. It is the natural generalisation of $\sigma$-models like \eqref{eq:NATDoriginal} or \eqref{eq:NATDdual}.

A first order parent action \cite{Klimcik1996c,Tyurin1996,Driezen2016}, generalising the one of abelian $T$-duality \eqref{eq:AbelianHamiltonianDensity}, for Poisson-Lie $\sigma$-models is a $\sigma$-model of fields $l$ taking value in a Drinfel'd double $\mathcal{D}$
\begin{align}
S &= \frac{1}{2} \int_{\partial B} \mathrm{d}^2 \sigma \ \left[ \braket{l^{-1}\partial_\sigma l , l^{-1} \partial_\tau l } - \braket{l^{-1}\partial_\sigma l , \hat{\mathcal{H}} (l^{-1} \partial_\sigma l)} \right] + \frac{1}{12} \int_B \braket{l^{-1} \mathrm{d} l \overset{\wedge}{,} [ l^{-1} \mathrm{d} l \overset{\wedge}{,} l^{-1}  \mathrm{d} l ] }, \label{eq:doubledPLaction}
\end{align}
where $\braket{ \ , \ }$ is the canonical metric on $\mathfrak{d}$. The operator $\hat{\mathcal{H}}$ represents a generalised metric
\begin{align}
\mathcal{H}_{AB} &\equiv  \mathcal{H}(T_A,T_B) \equiv \braket{T_A, \hat{\mathcal{H}} ( T_B) } = \left( \begin{array}{cc} G_0 - B_0 G_0^{-1} B_0 & B_0 G_0^{-1} \\ - G_0^{-1} B_0 & G_0^{-1} \end{array} \right). \nonumber
\end{align}
and is defined by constant symmetric resp. skewsymmetric $d\times d$-matrices $G_0$ resp. $B_0$ given in some basis $\{T_A\} = \{t_a ,\bar{t}^a\}$ of a Manin triple decomposition of $\mathfrak{d}$. As such \eqref{eq:doubledPLaction} is the natural generalisation of the toroidal first order action \eqref{eq:AbelianDoubledLag} with a few caveats:
\begin{itemize}
\item The polarisation term $\dot{X} \cdot P$ in the abelian case becomes a $WZW$-model like term.

\item The non-abelian nature of $\mathfrak{d}$ means that for some choice of Manin triple decomposition the decomposition of $l=\bar{g}g^{-1} \in \mathcal{D}$ for $\bar{g} \in \bar{\mathcal{G}}$ and $g \in \mathcal{G}$ will not result into a direct decomposition of $l^{-1} \partial_\sigma l$ (which would correspond to $(X^{\prime i} , P_i)$ in the abelian case), but instead we have:
\begin{align}
l^{-1} \mathrm{d} l &= \text{Ad}_g \left( - g^{-1} \mathrm{d} g + \bar{g}^{-1} \mathrm{d} \bar{g} \right) \label{eq:DecompositionMCdoubled} \\
&= - (g^{-1} \mathrm{d} g)^a \text{Ad}_g(t_a) + (\bar{g}^{-1} \mathrm{d} \bar{g})_a \text{Ad}_g(\bar{t}^a) \nonumber \\
&= \left( - (g^{-1} \mathrm{d} g)^a,(\bar{g}^{-1} \mathrm{d} \bar{g})_a \right) \left( \begin{array}{cc} \mathbb{1} & \\ \Pi(g) & \mathbb{1} \end{array} \right) \left( \begin{array}{cc} A(g) & \\ & (A^T)^{-1}(g) \end{array} \right) \left( \begin{array}{c} t^b \\ \bar{t}_b \end{array} \right) \nonumber
\end{align}
where the homogeneous Poisson structure $\Pi(g)$ arises according to definition \eqref{eq:PoissonHomogenousfromAd}.
\end{itemize}
What we call \textit{Poisson-Lie $\sigma$-model} in this paper is constructed as follows: Choose a Manin triple $\mathfrak{g} \oplus_{\mathfrak{d}} \mathfrak{g}^{\star}$ with a corresponding basis $\{t_a , \bar{t}^a\}$ and groups $\mathcal{G}$, $\bar{\mathcal{G}}$, and take a corresponding decomposition of $l$ as above $l = \bar{g} g^{-1}$. We put this choice of parameterisation of $\mathcal{D}$ into \eqref{eq:doubledPLaction}, then with knowledge of \eqref{eq:DecompositionMCdoubled} and help of the Polyakov-Wiegmann identity for the $WZW$-term we integrate out $\bar{g}$ and yield
\begin{equation}
S = \frac{1}{2} \int \mathrm{d}^2 \sigma \ (g^{-1} \partial_+ g)^a \left(\frac{1}{\frac{1}{G_0+B_0}+\Pi(g)}\right)_{ab} (g^{-1} \partial_- g)^b. \label{eq:PoissonLieSigmaModelOriginal}
\end{equation}
This model is a $\sigma$-model for $\mathcal{G}$-valued fields $g$. The bialgebra structure of the original doubled $\sigma$-model is encoded in the homogeneous Poisson structure $\Pi(g)$ of the form \eqref{eq:PoissonBivectorHomogeneous}, and the generalised metric finds itself in $E_0 = G_0 + B_0$. \pagebreak

The models, which we discussed before, belong to this class of $\sigma$-models for different choices of bialgebras:
\begin{itemize}
\item The \textit{toroidal $\sigma$-model} \eqref{eq:ActionTorus} is reproduced from \eqref{eq:PoissonLieSigmaModelOriginal} for $\mathfrak{d}$ being abelian.
\item We get the typical \textit{$\mathcal{G}$-isometric $\sigma$-model} \eqref{eq:NATDoriginal} for the so-called semi-abelian bialgebra 
\begin{equation}
\mathfrak{d} = \mathfrak{g} \oplus_{\mathfrak{d}} (\mathfrak{u}(1))^d . \nonumber
\end{equation}

\item The \textit{non-abelian $T$-dual} of the above \eqref{eq:NATDdual} corresponds then logically to 
\begin{equation}
\mathfrak{d} = (\mathfrak{u}(1))^d \oplus_{\mathfrak{d}} \mathfrak{g}. \nonumber
\end{equation}
Due to the abelian structure of the target space this Poisson structure of the form \eqref{eq:PoissonBivectorHomogeneous} is given by $\Pi_{ab}(x) = - {f^c}_{ab} x_c$.
\end{itemize}

\subsubsection*{Dirac structures}
The key data of the doubled $\sigma$-model \eqref{eq:doubledPLaction} is the bialgebra $\mathfrak{d}$ and the generalised metric $\mathcal{H}$. This data singles out a decomposition of $\mathfrak{d}$ into so-called Dirac structures, orthogonal subspaces w.r.t. to natural O$(d,d)$-metric $\braket{ \ , \ }$
\begin{equation}
\mathfrak{d} = \mathfrak{d}^+ \perp \mathfrak{d}^-. 
\end{equation}
Each choice of a non-degenerate $d\times d$-matrix $E_0 = G_0 +B_0$, with a metric $G_0$ and skewsymmetric $B_0$, chooses such a decomposition:
\begin{align}
S^\pm_a &= t_a \pm E_{0,ab}^\pm \bar{t}^b, \quad \text{with} \ E_0^+ = E_0^T, E_0^- = E_0 \\
\mathfrak{d}^\pm &= \text{span}\left( S^\pm_a \right).
\end{align}
This basis also block-diagonalises the canonical O$(d,d)$-metric
\begin{equation}
\braket{ \ , \ } = \ket{S^+_a} G_0^{ab} \bra{S^+_b} - \ket{S^-_a} G_0^{ab} \bra{S^-_b}
\end{equation}
So the effect of the generalised metric term in \eqref{eq:doubledPLaction} is to 'choose' a decomposition of $\mathfrak{d}$ into Dirac structures, which are the eigenspaces of $\hat{\mathcal{H}}$:
\begin{align}
\hat{\mathcal{H}}\big\vert_{\mathfrak{d}^\pm} &= \pm \braket{ \ , \ } \big\vert_{\mathfrak{d}^\pm}, \quad \hat{\mathcal{H}} = \ket{S^+_a} G_0^{ab} \bra{S^+_b} +  \ket{S^-_a} G_0^{ab} \bra{S^-_b}.
\end{align}
A crucial property of the (classical) doubled $\sigma$-model \eqref{eq:doubledPLaction} is, that the dynamics follows the constraint
\begin{align}
\braket{l^{-1} \partial_\pm l, \mathfrak{d}^\pm} = 0.
\end{align}
This relation highlights the role of the Dirac structures and was the starting point of the investigation of Poisson-Lie $T$-duality in \cite{Klimcik1995}, even before the doubled $\sigma$-model was discovered. For a mathematical treatment of Dirac structures and Courant algebroids in the context of Poisson-Lie $T$-duality see \cite{Severa2015}.

The choice of decomposition is of course independent of the basis choice of $\mathfrak{d}$, but the Dirac structure is \textit{non-manifestly} realised in the Poisson-Lie $\sigma$-model \eqref{eq:PoissonLieSigmaModelOriginal}, the explicit form of which will depend on a choice of basis. This is the key point in our analysis of Poisson-Lie $T$-duality.

\subsubsection*{Poisson-Lie $T$-duality}
Suppose we have another choice of Manin triple other than $\mathfrak{g}\oplus_{\mathfrak{d}}\mathfrak{g}^\star$ at our hand, the simplest choice of course being $\mathfrak{g}^\star \oplus_{\mathfrak{d}} \mathfrak{g}$. Let us choose a corresponding parametrisation of $\mathcal{D}$ in \eqref{eq:doubledPLaction} by $l=g \bar{g}^{-1}$ and integrate out $g$ (instead of $\bar{g}$ before). This gives rise to a classically equivalent $\sigma$-model
\begin{align}
S = \frac{1}{2} \int \mathrm{d}^2 \sigma (\bar{g}^{-1} \partial_+ \bar{g})_a \bar{E}^{ab}(\bar{g}) (\bar{g}^{-1} \partial_+ \bar{g})_b
\end{align}
with $\bar{E}^{-1}(g) = G_0 + B_0 + \bar{\Pi}$,
where $\bar{\Pi}$ is now the homogeneous Poisson structure on $\bar{\mathcal{G}}$, equivalent to the dual Lie algebra structure $\mathfrak{g}$. This is the Poisson-Lie $T$-dual of \eqref{eq:PoissonLieSigmaModelOriginal} and generalises the $R \leftrightarrow \frac{1}{R}$-behaviour from abelian $T$-duality by
\begin{align*}
E_0 \bar{E}_0^{-1} = \mathbb{1} \quad \text{and also} \quad E(e) \bar{E}(e) = \mathbb{1}.
\end{align*}
A Poisson-Lie $T$-duality group will consequently be the space of decompositions $l=\bar{h} h^{-1} \in \mathcal{D}$, where $h \in \mathcal{H}$, $\bar{h}\in \bar{H}$ for some $\mathcal{D} = H \Join \bar{H}$. On the Lie algebra level this is corresponds to the set of Manin triple decomposition of the bialgebra $\mathfrak{d}$ to $\mathcal{D}$. The task to explore this space is what we set about in the next chapter.

You can find generalisations of Poisson-Lie $T$-duality to coset spaces, open strings and supersymmetry \cite{Klimcik1996b,Klimcik1997,Sfetsos1996,Sfetsos1999}, a canonical analysis \cite{Klimcik1996c,Sfetsos1998,Stern1999,Cabrera2009} and studies of the dual models beyond the classical level \cite{ Alekseev1994,Tyurin1996,Sfetsos1998a,Parkhomenko1999,Sfetsos2009} in the literature.

\section{A non-abelian $T$-duality group}
\subsection{Definition}
\label{chap:DefNATD}
Motivated by the discussion in the last section about the construction of the Poisson-Lie $\sigma$-model \eqref{eq:PoissonLieSigmaModelOriginal} with a corresponding bialgebra $\mathfrak{d}$ from a doubled $\sigma$-model \eqref{eq:doubledPLaction} and following the first encounter with Poisson-Lie $T$-duality \cite{Klimcik1995}, a candidate for a non-abelian $T$-duality group\footnote{The investigation of this group, although it defined on basis of, what we called here, the Poisson-Lie $\sigma$-model, will turn out to contain isometric models like the principal chiral models and their non-abelian $T$-duals in many cases. As it will also turn out to be a direct generalisation of the abelian $T$-duality group O$(d,d)$, we decided in favour of the name 'non-abelian $T$-duality group' against 'Poisson-Lie $T$-duality group' here.} is
\begin{align}
\text{NATD group}(\mathfrak{d}) &= \left\{ \text{Manin triple decompositions of } \mathfrak{d} \right\}. \label{eq:NATDgroupNew} 
\end{align}
The elements of the NATD group as defined above is the set vector space automorphisms $\varphi$ of $\mathfrak{d}$, which
\begin{enumerate}
\item preserve the natural pairing $\braket{ \ | \ }$, so $\varphi \in $O$(d,d)$.
\item preserve the algebraic closure of $\mathfrak{g}$ and $\mathfrak{g}^\star$, i.e.
\begin{equation}
[ \varphi(\mathfrak{g}) , \varphi(\mathfrak{g}) ] \subset \varphi(\mathfrak{g}) \quad \text{and} \quad [ \varphi(\mathfrak{g}^\star) , \varphi(\mathfrak{g^\star}) ] \subset \varphi(\mathfrak{g^\star}).
\end{equation}
\end{enumerate}
In the case, where $\mathfrak{d}$ is abelian, \eqref{eq:NATDgroupNew} naturally becomes the O$(d,d)$ group of abelian $T$-duality. 

Let us emphasise that this group \eqref{eq:NATDgroupNew} is the modular space of Poisson-Lie $\sigma$-models corresponding to a bialgebra $\mathfrak{d}$ (and some given $G_0$, $B_0$). This does not imply that this group contains any (non-abelian) $T$-duality transformation, that we can think of. As a side note, we will find that condition 2 has to be partially relaxed in order to incorporate non-abelian $T$-dualities of principal chiral models with respect to subgroups. Nevertheless we will be content in the study of Poisson-Lie $\sigma$-models and thus focus on the above NATD group for most of this paper.

\textbf{Action on Poisson-Lie $\sigma$-model.} O$(d,d)$ basis transformations on the doubled $\sigma$-model \eqref{eq:doubledPLaction} result in different dual $\sigma$-models, because we project onto the same Dirac structure (which defines the model), but integrate out different d.o.f.s corresponding to $\varphi(\mathfrak{g}^\star)$. This similarity action is given by
\begin{itemize}
\item $(\mathfrak{d},\mathfrak{g}^\prime,\mathfrak{g}^{\prime \star} ) =  (\mathfrak{d},\varphi(\mathfrak{g}),\varphi(\mathfrak{g}^\star) )$, where again $\varphi$ is \textit{not a Lie algebra}, but only a \textit{vector space} automorphism. Generically there will be a change in algebraic structure.

\item standard O$(d,d)$-action on the generalised metric \eqref{eq:GeneralisedMetric} by the inverse $\varphi$
\begin{equation}
\mathcal{H}(G^\prime_0 + B^\prime_0) = \varphi^{-1} . \mathcal{H}(G_0 + B_0).
\end{equation}
\end{itemize}
So, in addition to transforming the background $G_0 + B_0$ as in abelian $T$-duality, we also need to account for the change in algebraic structure. The transformed $\sigma$-model looks like
\begin{align}
S = \frac{1}{2} \int \mathrm{d}^2 \sigma \ (g^{\prime -1} \partial_+ g^{\prime})^a \left(\frac{1}{\frac{1}{G^\prime_0 + B_0^\prime} + \Pi^\prime(g^\prime)}\right)_{ab} (g^{\prime -1} \partial_- g^{\prime})^b
\end{align}
where $g^{\prime}$ takes values in $\mathcal{G}^\prime \triangleleft \mathcal{D}$, which is the Lie group to $\varphi(\mathfrak{g})$, and $\Pi^\prime$ is the homogeneous Poisson bivector field on $\mathcal{G}^\prime$ corresponding to the transformed dual structure ${\overline{F}_c}^{ab}$.

\subsection{Standard subgroups}
\label{chap:NATDsubgroups}

In the subsequent literature the group \eqref{eq:NATDgroupNew} was studied systematically only for lower dimensional bialgebras and without physical interpretation of the transformations, following the Bianchi classification of three dimensional Lie algebras \cite{Jafarizadeh1999a,Unge2002,Snobl2002,Hlavaty2004,Hlavaty2007}. Now we want to understand some concrete structure of the NATD group apart from the original 'complete' factorised non-abelian $T$-duality transformations (called Poisson-Lie $T$-dualities so far) and give an explanation on the $\sigma$-model level or, if possible, a Buscher-like procedure. As the study of a generic $\varphi\in$ O$(d,d)$ is a little unhandy, we will make use of the standard decomposition of O$(d,d)$ into factories dualities and the three continuous subgroups: GL-transformations, $B$- and $\beta$-shifts. We look for the conditions such that these lie in the NATD group \eqref{eq:NATDgroupNew} and also for the meaning of these transformations on the level of the (undoubled) Poisson-Lie $\sigma$-model.

The study of the standard O$(d,d)$ subgroups should help to get physical understanding of this NATD group. The definition \eqref{eq:NATDgroupNew} will severely restrict the allowed factorised dualities, $B$-shifts and $\beta$-shifts. But it is by no means to expected that these subgroups generate all elements of \eqref{eq:NATDgroupNew}, and resultantly our investigation may only give a subgroup of \eqref{eq:NATDgroupNew}.

\textbf{Lie (bi)algebra automorphisms.} For example all the standard O$(d,d)$ subgroup transformations will turn out to be generically \textit{not} Lie algebra automorphisms. But Lie algebra automorphisms of $\mathfrak{d}$, that also preserve the O$(d,d)$-metric will also be part of the duality group, acting only on the background data $G_0$ and $B_0$. We will not consider these further.

\textbf{(Non-abelian $T$-duality) GL-transformations.} The simplest continuous subgroup of O$(d,d)$, general linear transformations GL$(d)$ of O$(d,d)$
\begin{equation}
\varphi_{GL} = \left( \begin{array}{cc} A^T & 0 \\ 0 & A^{-1} \end{array} \right), \quad \text{with } A \in \text{GL}(d)
\end{equation}
is clearly contained fully in the non-abelian $T$-duality group. It describes simultaneous basis changes of $\mathfrak{g}$ and $\mathfrak{g}^*$ preserving the O$(d,d)$-metric and also the algebraic closure conditions.

\subsubsection{Factorised non-abelian $T$-dualities}
Factorised dualities of O$(d,d)$ are the $\mathbb{Z}_2$-transformations corresponding to the maps
\begin{align}
\varphi_{f.d.}: \ t_\alpha &\mapsto t^\prime_\alpha = \bar{t}^\alpha, \ t_{\underline{\alpha}} \mapsto t^\prime_{\underline{\alpha}} = t_{\underline{\alpha}} \nonumber \\
\bar{t}^\alpha &\mapsto \bar{t}^{\prime\alpha} = t_\alpha, \ \bar{t}^{\underline{\alpha}} \mapsto \bar{t}^{\prime \underline{\alpha}} = \bar{t}^{\underline{\alpha}} \label{eq:phiNATDfactorisedDuality}
\end{align}
for $\alpha = 1,...,m$ and $\underline{\alpha} = m+1 , ... , d$, for an $m \leq d$.\footnote{This is generic as we can arrange any choice of the generators $\{ t_a \}$ with help of $GL$-transformations.} 
%The algebraic representation is 
%\begin{equation}
%\varphi_{f.d.} = \left( \begin{array}{cccc} 0_m & & \mathbb{1}_m &  \\  & \mathbb{1}_{d-m} & & 0_{d-m} \\  \mathbb{1}_m & & 0_m & \\ & 0_{d-m} & & \mathbb{1}_{d-m} \end{array} \right).
%\end{equation}
Following the definition \eqref{eq:NATDgroupNew} the $\varphi_{f.d.}$ are only NATD transformations, if the $\{t_a , \bar{t}^a \} = \{t_\alpha, t_{\underline{\alpha}} \ , \ \bar{t}^{\alpha},\bar{t}^{\underline{\alpha}} \}$ fulfil the following conditions:
\begin{equation}
{\overline{f}_{\underline{\gamma}}}^{\alpha \beta} = {\overline{f}_\gamma}^{\underline{\alpha} \underline{\beta}} = {f^{\underline{\gamma}}}_{\alpha \beta} = {f^\gamma}_{\underline{\alpha} \underline{\beta}} = 0 \label{eq:CondFactDualityStrong}
\end{equation}
This is equivalent to the decomposition of $\mathfrak{d}$
\begin{align}
\mathfrak{d}&= \left(\mathfrak{h} \oplus \mathfrak{m} \right) \oplus_{\mathfrak{d}} \left(\mathfrak{h}^\star \oplus \mathfrak{m}^\star \right) \label{eq:decompFactDualityStrong}
\\
\text{with} \qquad &{} [\mathfrak{h},\mathfrak{h}] \subset \mathfrak{h}, \qquad [\mathfrak{h}^\star , \mathfrak{h}^\star] \subset \mathfrak{h}^\star, \nonumber \\
&{} [\mathfrak{m},\mathfrak{m}] \subset \mathfrak{m}, \qquad [\mathfrak{m}^\star,\mathfrak{m}^\star] \subset \mathfrak{m}^\star, \nonumber \\
&{} [\mathfrak{h},\mathfrak{m}^\star] \subset \mathfrak{h}\oplus \mathfrak{m}^\star, \qquad [ \mathfrak{h}^\star , \mathfrak{m} ] \subset \mathfrak{h}^\star \oplus \mathfrak{m}, \nonumber
\end{align}
where $\mathfrak{h}$ resp. $\mathfrak{h}^\star$ is generated by $\{ t_\alpha \}$ resp. $\{ \bar{t}^\alpha \}$ and $\mathfrak{m}$ resp. $\mathfrak{m}^\star$ by $\{ t_{\underline{\alpha}} \}$ resp. $\{ \bar{t}^{\underline{\alpha}} \}$. Thus the factorised dualities act as
\begin{equation}
\left(\mathfrak{h} \oplus \mathfrak{m} \right) \oplus_{\mathfrak{d}} \left(\mathfrak{h}^\star \oplus \mathfrak{m}^\star \right)\quad \leftrightarrow \quad \left(\mathfrak{h}^\star \oplus \mathfrak{m} \right) \oplus_{\mathfrak{d}} \left(\mathfrak{h} \oplus \mathfrak{m}^\star \right) \label{eq:FactorisedDualNewBialgebra}
\end{equation}
on the bialgebra structure. The dual Poisson-Lie $\sigma$-model
\begin{equation}
L \quad \propto \quad j^{\prime a}_+ \left(\frac{1}{(E^{\prime}_0 )^{-1} + \Pi^\prime} \right)_{ab} j^{\prime b}
\end{equation}
consists of the Maurer-Cartan forms $j^\prime$ to the Lie group of the algebra $\mathfrak{h}^\star \oplus \mathfrak{m}$, the homogeneous Poisson structure $\Pi^\prime$ (corresponding to the new bialgebra \eqref{eq:FactorisedDualNewBialgebra}) and the transformed background $E^\prime_0$, given by the standard O$(d,d)$ action of $\varphi_{f,d}$ \eqref{eq:phiNATDfactorisedDuality} on the original $E_0$.

The conditions \eqref{eq:CondFactDualityStrong} seem to be very restrictive - even for the semi-abelian bialgebra \\ $\mathfrak{g}\oplus_{\mathfrak{d}} (\mathfrak{u}(1))^d$ with an abelian subalgebra $\mathfrak{h} \subset\mathfrak{g}$ these are not fulfilled in general, due to $[t_{\underline{\alpha}} , \bar{t}^\beta ] =  {f^\beta}_{\underline{\gamma}\underline{\beta}} \bar{t}^{\gamma} \notin \mathfrak{h}^\star \oplus \mathfrak{m} $ in general. In the following paragraph we study exactly this scenario - the non-abelian $T$-duality of the principal chiral model w.r.t. to a subgroup. This will indeed demonstrate that it cannot be put into the Poisson-Lie $\sigma$-model form \eqref{eq:PoissonLieSigmaModelOriginal}.

\subsubsection*{Subgroup non-abelian $T$-duality}
Consider the principal chiral model on a group $G$, which has a subgroup $H$ with Lie algebra $\mathfrak{h}$. We decompose the generators of $G$ correspondingly, $\{t_a\} = \{t_\alpha , t_{\underline{\alpha}} \}$, and the model by choosing $g = hm$ with $h \in H$ and some $m \in G$, so that
\begin{align}
g^{-1} \mathrm{d}g &= \text{Ad}_m^{-1} \left( h^{-1} \mathrm{d}h + \mathrm{d}m \ m^{-1} \right).
\end{align}
The action becomes
\begin{align}
S &\propto \int \text{Tr}\left( \left( h^{-1} \mathrm{d}h + \mathrm{d}m \  m^{-1} \right) \wedge \star \left( h^{-1} \mathrm{d}h + \mathrm{d}m \ m^{-1} \right) \right) \nonumber \\
&= \int \left\lbrace \text{Tr}\left( \left( A + \mathrm{d}m \ m^{-1} \right) \wedge \star \left( A + \mathrm{d}m \ m^{-1} \right) \right) - \bar{x}_\alpha \left(\mathrm{d}A^\alpha + [A \overset{\wedge}{,} A]^\alpha \right) \right\rbrace.
\end{align}
Integrating out the $\mathfrak{h}$-valued field $A$ yields
\begin{align}
\tilde{S} &\propto \int \mathrm{d}^2 \sigma \Big\lbrace \left(\partial_+ \bar{x}_\alpha + (\partial_+ m \ m^{-1} )^\sigma \kappa_{\sigma \alpha} \right) \frac{1}{\kappa_{\alpha \beta} - \bar{x}_\gamma {f^\gamma}_{\alpha \beta}} \left( \partial_- \bar{x}_\beta - \kappa_{\beta \tau} (\partial_- m \ m^{-1})^{\tau} \right)  \nonumber \\
&{} \qquad  + \text{Tr} \left( (\partial_+ m \ m^{-1}) \ (\partial_- m \ m^{-1}) \right) \Big\rbrace . \label{eq:NATDsubgroupAction}
\end{align}
The equations of motion can be expressed in the following form
\begin{align}
\mathrm{d} \bar{J}_a + \frac{1}{2} {Q_a}^{b c} \bar{J}_b \wedge \bar{J}_c = 0
\end{align}
with the Poisson structure $\Pi_{\alpha \beta} = - \bar{x}_\gamma {f^\gamma}_{\alpha \beta}$ and the current\footnote{Superficially it looks as if this would have increased the degrees of freedom, but the variation of \eqref{eq:NATDsubgroupAction} with respect to $(\partial m \ m^{-1})^\alpha$ vanishes by the equations of motion for $\bar{x}_\alpha$.}
\begin{align}
\bar{J}_{\pm} = \pm \left(\frac{1}{\mathbb{1} \pm \Pi}\right)_{\alpha \beta} \left( \partial_\pm \bar{x}^\beta \pm (\partial_\pm m \ m^{-1}) \right) \bar{t}^{\prime \alpha} \pm \delta_{\underline{\alpha} \underline{\beta}}(\partial_\pm m \ m^{-1})^{\underline{\alpha}} \bar{t}^{\prime \underline{\beta}}
\end{align}
and the new generators of the transformation \eqref{eq:phiNATDfactorisedDuality}, where the $\{ t_\alpha \}$ generate a subalgebra:
\begin{small}
\begin{align}
[t^\prime_\alpha , t^\prime_\beta] &= 0 \equiv {F^c}_{\alpha \beta} t^\prime_c + H_{\alpha \beta c} \bar{t}^{\prime c}, &{} [\bar{t}^{\prime \alpha},\bar{t}^{\prime \beta}] = {f^\gamma}_{\alpha \beta} \bar{t}^{\prime \gamma} \equiv {Q_\gamma}^{\alpha \beta} \bar{t}^{\prime \gamma} + R^{\alpha \beta c} {t}^{\prime}_c \nonumber \\
[ t^\prime_\alpha, t^\prime_{\underline{\beta}} ] &= {f^\alpha}_{\gamma \underline{\beta}} t^\prime_\gamma +  {f^\alpha}_{\underline{\gamma} \underline{\beta}} \bar{t}^{\prime \underline{\gamma}} \equiv {F^\gamma}_{\alpha \underline{\beta}} t^\prime_\gamma + H_{\alpha \underline{\beta} \underline{\gamma} } \bar{t}^{\prime \underline{\gamma}}, &{} [\bar{t}^{\prime \alpha}, \bar{t}^{\prime \underline{\beta}}] = {f^{\underline{\beta}}}_{\underline{\gamma} \alpha} \bar{t}^{\prime \underline{\gamma}} \equiv {Q_{\underline{\gamma}}}^{\alpha \underline{\beta}} \bar{t}^{\prime \underline{\gamma}} + R^{\alpha \beta c} t^\prime_c \nonumber \\
[ t^\prime_{\underline{\alpha}}, t^\prime_{\underline{\beta}} ] &= {f^{\underline{\gamma}}}_{\underline{\alpha} \underline{\beta}} t^\prime_{\underline{\gamma}} +  {f^\gamma}_{\underline{\alpha} \underline{\beta}} \bar{t}^{\prime \gamma} \equiv {F^{\underline{\gamma}}}_{\underline{\alpha} \underline{\beta}} t^\prime_{\underline \gamma} + H_{\underline{\alpha} \underline{\beta} \gamma } \bar{t}^{\prime {\gamma}}, &{} [\bar{t}^{\prime \underline{\alpha}} , \bar{t}^{\prime \underline{\beta}}] = 0 \equiv {Q_c}^{\underline{\alpha} \underline{\beta}} \bar{t}^{\prime c} + R^{\underline{\alpha} \underline{\beta} c} t^\prime_c \nonumber \\
[t^\prime_\alpha,\bar{t}^{\prime \beta}] &= {f^\alpha}_{\beta \gamma} t^\prime_\gamma + {f^\alpha}_{\beta \underline{\gamma}} \bar{t}^{\prime \underline{\gamma}} \equiv  {F^\beta}_{\alpha \underline{\gamma}} \bar{t}^{\prime \underline{\gamma}} + {Q_\alpha}^{\beta \gamma} t^\prime_\gamma, &{} [t^\prime_\alpha, \bar{t}^{\prime \underline{\beta}}] = 0 \equiv {F^{\underline{\beta}}}_{c \alpha} \bar{t}^{\prime c} + {Q_\alpha}^{\underline{\beta} c} t^\prime_c \nonumber \\
[t^\prime_{\underline{\alpha}},\bar{t}^{\prime \beta}] &= {f^\gamma}_{\underline{\alpha} \beta} t^\prime_\gamma + {f^{\underline{\gamma}}}_{\underline{\alpha} \beta} \bar{t}^{\prime \gamma} \equiv {F^\beta}_{\gamma \underline{\alpha}} \bar{t}^{\prime \gamma} + {Q_{\underline{\alpha}}}^{\beta \underline{\gamma}} t^\prime_\gamma, &{} [t^\prime_{\underline{\alpha}}, \bar{t}^{\prime \underline{\beta}} ] = {f^{\underline{\beta}}}_{\gamma \underline{\alpha}} t^\prime_\gamma + {f^{\underline{\beta}}}_{\underline{\gamma} \underline{\alpha}} \bar{t}^{\prime \underline{\gamma}} = {F^{\underline{\beta}}}_{\underline{\gamma} \underline{\alpha}} \bar{t}^{\prime \underline{\gamma}} + {Q_{\underline{\alpha}}}^{\underline{\beta} \gamma} t^\prime_\gamma , \label{eq:NATDsubgroupAlgebra} 
\end{align}
\end{small}where we organised the resulting structure constants in the conventions of generalised fluxes.\footnote{Because we start with a semi-abelian bialgebra and the ${f^c}_{ab}$ fulfil the Jacobi identity, the coefficients $H$, $F$, $Q$ and $R$ fulfil the standard Bianchi identities of non-geometric fluxes \cite{Shelton2005}. We will comment further on this topic in section \ref{chap:GeneralisedFluxes}.} So, formally \eqref{eq:NATDsubgroupAction} looks like a Poisson-Lie $\sigma$-model, but the current
\begin{equation}
\left(\partial_\pm \bar{x}_\alpha \pm (\partial_\pm m \ m^{-1})^\sigma \kappa_{\sigma \alpha} \right) \bar{t}^\alpha + (\partial_\pm m \ m^{-1})^{\underline{\alpha}} t_{\underline{\alpha}} \in \mathfrak{h}^\star \oplus \mathfrak{m} \label{eq:subgroupNATDcurrent}
\end{equation}
is not the Maurer-Cartan form of a group, because $\mathfrak{h}^\star \oplus \mathfrak{m}$ is not closed under the Lie bracket generically. As long as ${f^\alpha}_{\underline{\beta} \underline{\gamma}}$ (the only non vanishing component of $H$ in \eqref{eq:NATDsubgroupAlgebra}) does not vanish, it does not seem possible to arrange the Bianchi identities of \eqref{eq:subgroupNATDcurrent} into a zero curvature form, which would required in order for the subgroup non-abelian $T$-dual model to be of the Poisson-Lie $\sigma$-model form - this agrees with \eqref{eq:NATDgroupNew}.

\subsubsection*{A modified definition of a NATD group}
Motivated by the above considerations, let us give a refined version for the definition of the NATD group
\begin{align}
\text{mod. NATD group} \left(\mathfrak{d}\right) &= \left\lbrace \text{Manin pair decompositions of } \mathfrak{d}\right\rbrace \label{eq:NATDgroupWeak} \\
&\simeq \left\lbrace (\varphi: \ \mathfrak{d}\rightarrow \mathfrak{d}) \in\text{O}(d,d) : \ [\varphi(\mathfrak{g}^\star) , \varphi(\mathfrak{g}^\star)] \subset \varphi(\mathfrak{g}^\star) \right\rbrace, \nonumber 
\end{align}
which goes beyond the notion of Poisson-Lie $\sigma$-models, but includes the previous case.\footnote{A Manin pair $(\mathfrak{d},\mathfrak{g}^\star)$ is a pair consisting of a $2d$-dimensional Lie algebra $\mathfrak{d}$ admitting an O$(d,d)$-metric and a Lagrangian subalgebra of $\mathfrak{d}$, here denoted by $\mathfrak{g}^\star$.} From the perspective of the construction from a doubled $\sigma$-model \eqref{eq:doubledPLaction}, the above scenario is plausible, because for integrating out the degrees of freedom consistency requires only, that $\mathfrak{g}^\star$ is a subalgebra (resp. $\bar{\mathcal{G}}$ a subgroup), but not $\mathfrak{g}$. The refined conditions in comparison to \eqref{eq:CondFactDualityStrong} for factorised dualities were already stated in \cite{Hoare2017} with slight differences\footnote{In \cite{Hoare2017} the authors required that for subgroup Poisson-Lie $T$-duality the dual flatness condition should decompose fully 
\begin{equation}
\mathrm{d} \bar{j}_\alpha + \frac{1}{2} {\barf_\alpha}^{bc} \bar{j}_b \bar{j}_c \equiv \mathrm{d} \bar{j}_\alpha + \frac{1}{2} {\barf_\alpha}^{\beta \gamma} \bar{j}_\beta \wedge \bar{j}_\gamma = 0. \label{eq:ConditionSeibold}
\end{equation}
But the factorised Poisson-Lie $T$-duality map $\varphi$ acting on the currents $j=g^{-1} \mathrm{d}g$ and $\bar{j}_{\pm,a} = \pm \left( \frac{1}{E_0^{-1} \pm \Pi}\right)_{ab} j^b_\pm$ is more complicated than $j^\alpha \leftrightarrow \bar{j}_\alpha$ and $E_0 \leftrightarrow E_0^{-1}$ in the abelian case - it consists also of exchanging $\Pi \rightarrow \Pi^\prime$. Our analysis shows, that condition \eqref{eq:ConditionSeibold} is not required for consistency of Poisson-Lie $T$-duality}:
\begin{align}
&{} \left(\mathfrak{h} \oplus \mathfrak{m} \right) \oplus_{\mathfrak{d}} \left(\mathfrak{h}^\star \oplus \mathfrak{m}^\star \right) \label{eq:decompFactDualityWeak}
\\
\text{with} \quad &{} [\mathfrak{h},\mathfrak{h}] \subset \mathfrak{h}, \quad [\mathfrak{m}^\star,\mathfrak{m}^\star] \subset \mathfrak{m}^\star\quad \text{and} \quad [\mathfrak{h},\mathfrak{m}^\star] \subset \mathfrak{h}\oplus \mathfrak{m}^\star, \nonumber
\end{align}
Of course the setting of bialgebras resp. Drinfel'd doubles, Poisson-Lie $\sigma$-models and Poisson-Lie $T$-duality is very narrow. A consistent treatment of the modified definition \eqref{eq:NATDgroupWeak} would require a different setting, i.e. one, which is \textit{not} based on a bialgebra $\mathfrak{d}$, but on any even dimensional Lie algebra, which admits an O$(d,d)$-invariant metric and has one maximally isotropic subalgebra. This would include very different settings, of course bialgebras but e.g. also the setup discussed in \cite{Hassler2016}, with a symmetric space decomposition $\mathfrak{d} = \mathfrak{m}^{(0)} \oplus \mathfrak{m}^{(1)}$, where $\mathfrak{m}^{(0)}$ is an isotropic subspace and -algebra w.r.t. to the O$(d,d)$ metric and $\mathfrak{m}^{(1)}$ is complementary isotropic subspace, but fulfils $[\mathfrak{m}^{(1)},\mathfrak{m}^{(1)}] = \mathfrak{m}^{(0)}$ and thus does not close.

Nevertheless we will continue to work with the more restrictive Manin triple definition \eqref{eq:NATDgroupNew}, as it gives already some interesting insights in the subgroups in the component connected to identity of the duality group \eqref{eq:NATDgroupNew}, which we will study in the following.

\subsubsection{Non-abelian $T$-duality $B$-shifts}
Let us come to the $B$-shift and $\beta$-shift subgroups of the NATD group, which have not been considered so far in the literature. In context of abelian $T$-duality $B$-shifts correspond to gauge transformations of the Kalb-Ramond field $B$, leaving the $\mathbf{H}$-flux, $\mathbf{H} = \mathrm{d}B$, invariant. The expectation is that this behaviour generalises to the $\mathbf{H}$-flux of the Poisson-Lie $\sigma$-model \eqref{eq:PoissonLieSigmaModelOriginal} and the $B$-shifts of the NATD group \eqref{eq:NATDgroupNew} (from now on NATD $B$-shifts). We will see hints for this in section \ref{chap:GeneralisedFluxes}. 

In this section we are going to discuss, how the NATD $B$-shift looks and how it acts on the Poisson-Lie $\sigma$-model. $B$-shifts in O$(d,d)$ are of the form
\begin{align}
\varphi_B &= \left( \begin{array}{cc} \mathbb{1} & \sigma_{ab} \\ 0 & \mathbb{1} \end{array} \right) \label{eq:B-shift}
\end{align}
with a skewsymmetric $d \times d$-matrix $\sigma_{ab}$. The transformed algebra relations are
\begin{align}
[{t}^\prime_a , \bar{t}^\prime_b] &= {F^c}_{ab} {t}^{\prime}_c + H_{abc} \bar{t}^{\prime c} \label{eq:B-shiftClosureg} \\
\text{with} \quad {F^c}_{ab} &= {f^c}_{ab} + \sigma_{k(a} {\barf_{b)}}^{kc} \quad \text{and} \quad H_{abc} = \sigma_{(a|d} \sigma_{|b|e} {\barf_{|c)}}^{de} - \sigma_{k(a} {f^k}_{bc)} \overset{!}{=} 0, \label{eq:B-shiftCondition}
\end{align}
by imposing algebraic closure on $\varphi_B(\mathfrak{g})$. The only relevant Jacobi identities, we need to check for these new algebra relations \eqref{eq:B-shiftClosureg} are:
\begin{align}
{F^k}_{(ab} {F^d}_{c)k} &= {f^k}_{ab} {f^d}_{ck} - \sigma_{ma} \sigma_{nb} {\barf_k}^{(mn} {\barf_c}^{d)k} + {\barf_c}^{dk} H_{abk} \nonumber \\
&{}\quad + \sigma_{cm} \left( {f^k}_{ab} {\barf_k}^{dm} - {f^{(d}}_{k(a} {\barf_{b)}}^{m)k} \right) + \left(\text{c. p. of }(abc)\right) \label{eq:B-shiftJacobiCheck}\\
{F^k}_{ab} {\barf_k}^{mn} - {F^{(m}}_{k(a} {\barf_{b)}}^{n)k} &= {f^k}_{ab} {\barf_k}^{mn} - {f^{(d}}_{k(a} {\barf_{b)}}^{n)k} + \sigma_{l(a} {\barf_{b)}}^{l(k} {\barf_k}^{mn)} \nonumber
\end{align}
which vanish due to Jacobi identities of the original bialgebra and the condition \eqref{eq:B-shiftCondition}, $H \equiv 0$. This condition is the crucial requirement of a NATD $B$-shift in comparison to the abelian case, where it is trivially fulfilled. Let us distinguish three cases to understand it better:
\begin{enumerate}
\item \underline{$\mathfrak{g}^\star$ is abelian: ${\barf_c}^{ab} \equiv 0$}

Then $0\equiv H_{abc} = - \sigma_{k(a} {f^k}_{bc)}$, means that $\sigma_{ab} (g^{-1} \mathrm{d} g)^a \wedge (g^{-1} \mathrm{d}g)^b $ is a  closed 2-form on $\mathcal{G}$. In this case the NATD $B$-shift simply adds of a 2-cocycle term to the Lagrangian:
\begin{align}
S &\propto \int \mathrm{d}^2\sigma (g^{-1} \partial_+ g)^a \left[ G_0 + B_0 + \sigma \right]_{ab} (g^{-1} \partial_- g)^b,
\end{align}
which is a gauge transformation of the $\mathbf{H}$-flux. Later we will argue that this is a generic feature also in the generic case.

\item \underline{$\mathfrak{g}$ is abelian: ${f^c}_{ab} \equiv 0$}

$\sigma_{ab} \bar{t}^a \wedge \bar{t}^b$ is a solution of the classical Yang-Baxter equation on $\mathfrak{g}^\star$. In this case a $\sigma$-model interpretation is possible in the dual picture - there $B$-shifts will be $\beta$-shifts of an isometric model. We will show in the next paragraph on NATD $\beta$-shifts, that these are indeed easier to understand in this specific case and that we can employ a generalised Buscher procedure there. 

\item \underline{generic case}:

Generically $\sigma_{ab}$ will be neither a 2-cocycle on $\mathfrak{g}$, nor a solution to the classical Yang-Baxter equation on $\mathfrak{g}^\star$. \eqref{eq:B-shiftCondition} says that the failures for both cancel each other out. We can also view \eqref{eq:B-shiftCondition} as 2-cocycle condition of $\sigma_{ab}$ w.r.t. the \textit{new} structure constants in \eqref{eq:B-shiftCondition}
\begin{align}
H_{abc} = - \sigma_{k(a} {F^k}_{bc)} = 0.
\end{align}
If we restrict to start with ${f^c}_{ab}$ being a 1-coboundary algebra to ${\barf_c}^{ab}$ with ${f^c}_{ab}= - {\bar{f}_{(a}}^{bc} \tau_{b)d}$ for some $\tau= \tau_{ab} \bar{t}^a \wedge \bar{t}^b$ and make the ansatz $\sigma_{ab} = \tau^\prime_{ab} - \tau_{ab}$, condition \eqref{eq:B-shiftCondition} becomes
\begin{equation}
{\barf_e}^{cd} \left[ \tau^{\prime}_{ac} \tau^{\prime}_{bd} - \tau_{ac} \tau_{bd} \right] + \text{c.p. of }(abe)= 0 \label{eq:B-ShiftCondition2}
\end{equation}
which is satisfied if $\tau^\prime$ and $\tau$ fulfil \textit{the same} Yang-Baxter \textit{like} equation. In this case we can understand a NATD $B$-shift as
\begin{itemize}
\item exchanging the 1-coboundary bialgebra structures ${f^c}_{ab} \leftrightarrow {F^c}_{ab} $ that fit to $\mathfrak{g}^\star$, which is unaffected by the $B$-shift. This will also change $\Pi \rightarrow \tilde{\Pi}$ with $\tilde{\Pi}$ being of the standard form \eqref{eq:PoissonBivectorHomogeneous} corresponding to the dual structure constants ${\barf_c}^{ab}$ but now on a new group with structure constants ${F^c}_{ab}$.

\item standard action of $\varphi_B$ on $E_0 = G_0 + B_0$.
\end{itemize}
\end{enumerate}
A generic Buscher-like procedure or some other action on the (non-doubled) Lagrangian level, which reproduces the NATD $B$-shift action, has not been found yet, but it is not necessarily expected to exist, as there is also none for the factorised dualities. The justification for these transformation thus lies in the common origin in the same doubled $\sigma$-model \eqref{eq:doubledPLaction}.

\subsubsection{Non-abelian $T$-duality $\beta$-shifts}
From the perspective of the definition of the NATD group \eqref{eq:NATDgroupNew}, $\beta$-shifts are exactly conjugate to the previous encountered NATD $B$-shifts. On the other hand the formulation of the Poisson-Lie $\sigma$-model is not duality symmetric, so NATD $\beta$-shifts and their action on Poisson-Lie $\sigma$-model deserve some attention on their own. $\beta$-shifts in O$(d,d)$ are of the form
\begin{align}
\varphi_\beta &= \left( \begin{array}{cc} \mathbb{1} & 0 \\ r^{ab} & \mathbb{1} \end{array} \right) \label{eq:beta-shift}
\end{align}
with a skewsymmetric $d \times d$-matrix $r^{ab}$. The transformed algebra relations are
\begin{align}
[\bar{t}^{\prime a} , \bar{t}^{\prime b}] &= {\overline{F}_c}^{ab} \bar{t}^{\prime c} + R^{abc} t^\prime_c \label{eq:beta-shiftClosuregStar} \\
\text{with} \quad {\overline{F}_c}^{ab} &= {\barf_c}^{ab} + r^{k(a} {f^{b)}}_{kc} \quad \text{and} \quad R^{abc} = r^{(a|d} r^{|b|e} {f^{|c)}}_{de} - r^{k(a} {\barf_k}^{bc)} \overset{!}{=} 0, \label{eq:beta-shiftCondition}
\end{align}
Also the solutions to the closure condition are basically the same as those in the $B$-shift case. In the dual picture (understand a NATD $\beta$-transformation as the sequence \textit{NATD factorised duality} - \textit{NATD $B$-shift} - \textit{NATD factorised duality}) the interpretation is exactly the same as NATD $B$-shifts.
\begin{enumerate}
\item \underline{$\mathfrak{g}$ is abelian: ${f^c}_{ab} \equiv 0$}

$r = r^{ab} t_a \wedge t_b$ is a symplectic 2-form on $\mathfrak{g}^\star$. In this case the NATD $\beta$-shift is indeed easiest understood in the dual picture, where it is simply a NATD $B$-shift with the intuitive $\sigma$-model interpretation as discussed in the previous paragraph.

\item \underline{$\mathfrak{g}^\star$ is abelian: ${\barf_c}^{ab} \equiv 0$}

Then $r = r^{ab} t_a \wedge t_b$ is a solution of the classical Yang-Baxter equation on $\mathfrak{g}\otimes \mathfrak{g}$. The resulting $\sigma$-model will be discuss below. 

\item \underline{generic case:} 

If both $\mathfrak{g}$ and $\mathfrak{g}^\star$ are non-trivial, then we use \eqref{eq:beta-shiftCondition}. Let us restrict to 1-coboundary bialgebras with ${\barf_c}^{ab}= - {f^{(a}}_{bc} s^{b)d}$ for some $s=s^{ab} t_a \wedge t_b$. With the ansatz $r^{ab} = s^{\prime ab} - s^{ab}$ condition \eqref{eq:beta-shiftCondition} becomes
\begin{equation}
{f^e}_{cd} \left[ s^{\prime ac} s^{\prime bd} - s^{ ac} s^{ bd} \right] + \text{c.p. of }(abe)= 0, \label{eq:beta-ShiftCondition2}
\end{equation}
which is satisfied if $s^\prime$ and $s$ fulfil \textit{the same} Yang-Baxter \textit{like} equation. NATD $\beta$-shifts switch between different choices of dual Lie algebra for a given ${f^c}_{ab}$, in a way that $r$ corresponds to a 2-cocycle on the new dual algebra
\begin{equation}
r^{k(a} {\overline{F}_k}^{bc)} = 0.
\end{equation}
This incorporates the cases, where we can '$\beta$-untwist' to standard $\mathcal{G}$-isometric $\sigma$-models. 

Exactly dual to the generic NATD $B$-shift case, these kind of NATD $\beta$-transformations can be thought of as connecting Poisson-Lie $\sigma$-models for the same quasi-isometry algebra $\mathfrak{g}$ but different dual structure $\mathfrak{g}^\star$ connected by \eqref{eq:beta-ShiftCondition2} fulfilling condition \eqref{eq:beta-shiftCondition}. In general the NATD $\beta$-shifted Poisson-Lie $\sigma$-model is given by
\begin{equation}
S = \frac{1}{2} \int \mathrm{d}^2 \sigma \ (g^{-1} \partial_+ g)^a \left(\frac{1}{\frac{1}{G_0+B_0}-r+\Pi^\prime(g)}\right)_{ab} (g^{-1} \partial_- g)^b. 
\end{equation}
\end{enumerate}
\textbf{$\sigma$-model interpretation via generalised Buscher procedure}. Generically, exactly as for the NATD $B$-shifts or factorised dualities, a derivation of NATD $\beta$-shifts only on the (non-doubled) Lagrangian level is not available. But in the semi-abelian case, the $\mathcal{G}$-isometric model \eqref{eq:NATDoriginal}, it is possible and mediated by the non-abelian generalisation of the 'generalised Buscher procedure' (see section \ref{chap:betaTransformationAbelian}). It was introduced already in \cite{Hoare2016} to show for certain examples on AdS${}_5$, that homogeneous Yang-Baxter deformations are non-abelian $T$-duality transformations, and explained further in \cite{Hoare2016c}. We explain this generalised Buscher procedure in generality here
\begin{enumerate}
\item Start with a Lie group $\mathcal{G}$ and consider the following $\sigma$-model for group $\mathcal{G}$
\begin{equation}
S = \frac{1}{2} \int \left(g^{-1} \mathrm{d} g \right)^a \wedge \left(G_0 \star \ + B_0 \right)_{ab} \left(g^{-1} \mathrm{d} g \right)^b, \label{eq:ActionSemiAbelianDD}
\end{equation}
with constant metric $G_0$ and $B_0$.
\item Given a 2-cocycle $\omega$ on $\mathfrak{g}$ we define a central extension $\mathfrak{e}$ of $\mathfrak{g}$ by a central element $Z$ with the new bracket $[ \ , \ ]^\prime$ by
\begin{equation}
[t_a,t_b] = {f^c}_{ab} t_c \quad \rightarrow \quad [t_a,t_b]^\prime = [t_a,t_b] + \omega_{ab} Z
\end{equation}
and the field strength $F^\prime$ of an $\mathfrak{e}$-valued gauge field $A^\prime=A^a t_a + C Z$ by
\begin{align}
F^\prime &= \mathrm{d}A^\prime - [A^\prime\overset{\wedge}{,}A^\prime]^\prime \\
F^{\prime,a} &= F^a = \mathrm{d}A^a - [A\overset{\wedge}{,}A]^a \quad \text{and} \quad F^{\prime,Z} = \mathrm{d} C - \omega_{ab}A^a \wedge A^b. \nonumber
\end{align}
\item Again on a symplectic leaf of the 2-cocycle $\omega$, which actually defines a subalgebra\footnote{A closed Chevalley-Eilenburg 2-cocycle defines a so-called quasi-Frobenius subalgebra, which is exactly the space, where the 2-cocycle is non-degenerate. \cite{Borsato2016}}, this defines a Poisson structure $\Pi(g)$ 
\begin{equation}
\omega = \omega_{ab} \bar{t}^a \wedge \bar{t}^b = \Pi^{-1}_{\alpha \beta} \bar{t}^{\alpha} \wedge \bar{t}^{\beta}.
\end{equation}
Gauging $\left(g^{-1} \mathrm{d} g \right)^a \mapsto A^a$ but fixing the field strength $F^\prime$ to be zero, instead of $F$, via adding the Lagrangian multiplier term
\begin{equation}
\mathcal{L}_{Lag.mult.} \propto - Y_s F^s = \bar{X}_a \wedge A^a + Y
\end{equation}
and integrating out $C$ and the Lagrangian multipliers $Y_s = (\bar{X}_a,Y)$ leaves, similarly to earlier calculations, the $\sigma$-model
\begin{equation}
S = \int \mathrm{d}^2 \sigma \ \left(g^{-1} \partial_+ g \right)^a \left(g^{-1} \partial_- g \right)^b \left(\frac{1}{\frac{1}{G_0+B_0} + \Pi}\right)_{ab}.
\end{equation}
\end{enumerate}
As the Poisson bivector $\Pi$ here should be invertible on a symplectic leaf $T\mathcal{G}$ and thus non-vanishing everywhere on $\mathcal{G}$, it can be only of the forms
\begin{equation}
\Pi_R^{ab}(g) = r^{ab} \quad \text{or} \quad \Pi^{ab}_L\left(g = \exp(x^a t_a) \right) = r^{ab} - r^{k(a} {f^{b)}}_{kc} x^c + ...
\end{equation}
as discussed in section \ref{chap:BialgebraDef}. In the sense of our definition of the action on \eqref{eq:ActionSemiAbelianDD} of the NATD group in section \ref{chap:DefNATD} only the latter has the striven for form\footnote{In fact both versions make sense, as
\begin{equation}
(g^{-1} \partial_+ g)^a \left(\frac{1}{E_0^{-1} + \Pi_L }\right)_{ab} (g^{-1} \partial_- g)^b = (\partial_+ g \ g^{-1} )^a \left(\frac{1}{A^{T} E_0^{-1} A + \Pi_R }\right)_{ab} (\partial_- g \ g^{-1} )^b
\end{equation}
The constant $\Pi_R$ corresponds to the $\beta$-shift (and a GL-transformation by $A(g)$ of the inner automorphism corresponding to the adjoint action on $g$ acting on $E_0$) corresponding to the right isometries $G_R$ of the principal chiral model. This is well known in contexts of Yang-Baxter deformations, where $E_0$ is ad-invariant. We will come back to this later.}, which agrees with the one of a NATD $\beta$-shifts by $-r$.

Our conventions on Lie algebra cohomology and the connection of 2-cocycles to central extensions are very briefly reviewed in appendix \ref{chap:Cohomology}.

\section{Generalised fluxes}
\label{chap:GeneralisedFluxes}
\subsection{Review and definition}
Starting with some (geometric) background, $T$-dualities can result in backgrounds with non-geometric features - namely non-commutativity and non-associativity - characterised by so-called non-geometric fluxes. Double field theory and generalised geometry as its mathematical tool helped to understand the web of connected backgrounds and also the nature of the non-geometric backgrounds better. Motivated by the fact, that the description of Poisson-Lie $\sigma$-model and bialgebras resembles generalised geometry, we study the Poisson-Lie $\sigma$-models as realisation of a (classical) string in a non-trivial generalises flux background. For this we review these and give their definitions and Bianchi identities.

\textbf{Motivation.} Starting with a three-torus with constant $\mathbf{H}$-flux, $\mathbf{H} = h \mathrm{d}x\wedge \mathrm{d}y \wedge \mathrm{d}z$, we can successively perform (factorised) abelian $T$-duality transformations along the two isometries\footnote{Due to $B_{xy}=h z$, for example as one possible choice for $x,y,z$ as coordinates on $T^3$, the presence of $H$-flux breaks one of the isometries of $T^3$.} of the background, and also a formal $T$-duality along the third direction. The resulting backgrounds are usually summarised in the diagram \cite{Blumenhagen2011}
\begin{equation}
\mathbf{H}_{xyz} \overset{T_x}{\longleftrightarrow} {\mathbf{f}^x}_{yz} \overset{T_y}{\longleftrightarrow} {\mathbf{Q}_z}^{xy} \overset{T_z}{\longleftrightarrow} \mathbf{R}^{xyz}. \label{eq:NonGeometricFluxChain}
\end{equation}
Let us briefly sketch the nature of these backgrounds:
\begin{itemize}
\item $\mathbf{H}$-flux background: The original $T^3$ with constant $\mathbf{H}$-flux.
\item $\mathbf{f}$-flux background: Applying the Buscher rules gives a new geometry, often called 'twisted torus', described by a Heisenberg algebra.

\item $\mathbf{Q}$-flux backgrounds, also called $T$-folds, are locally geometric spaces. In the above setting we have locally a $T^2$-fibration along the base $S^1$. Globally, whilst transporting the fibre around the base once, we need to glue the patches together with help of a $T$-duality transformation, not a diffeomorphism.

Another point of view on these spaces is, that the coordinates describing them do not commute $[x,y] \propto {\mathbf{Q}_z}^{xy} w_z$, where $w_z$ is the winding of the string around the $z$-cycle.

\item An $\mathbf{R}$-flux background is a locally and globally non-geometric 'space'. The coordinates do not associate anymore. Here we have $[[x,y],z] + \ \text{c.p.} \ \propto \mathbf{R}^{xyz}$.

\end{itemize}
More details on non-commutativity and non-associativity due to $\mathbf{Q}$- and $\mathbf{R}$-flux can be found in \cite{Blumenhagen2011a,Lust2010,Blumenhagen2011,Condeescu2012,Andriot2012a,Mylonas2012,Bakas2013}.

\textbf{Definition.} In an NS-NS background, given by a metric $G$ and a Kalb-Ramond field $B$, neglecting again the dilaton, the generalised fluxes are defined in a non-holonomic basis $t_a = {e_a}^i \partial_i \equiv "\partial_a"$ by \cite{Grana2009}
\begin{align}
\mathbf{H}_{abc} &= \partial_{(a} B_{bc)} + {f^d}_{(ab} B_{c)d} \nonumber \\
{\mathbf{f}^c}_{ab} &= {e^c}_j \left({e_a}^i \partial_i {e_b}^j - {e_b}^i \partial_i {e_a}^j \right) = {f^c}_{ab} \nonumber \\
{\mathbf{Q}_c}^{ab} &= \partial_c \beta^{ab} + {f^{(a}}_{mc} \beta^{b)m} \nonumber \\
\mathbf{R}^{abc} &= - \beta^{m(a} \partial_m \beta^{bc)} + {f^{(a}}_{mn} \beta^{b|m} \beta^{|c)n}. \label{eq:GeneralisedFluxesDef}
\end{align}
The fluxes cannot be turned on and off independently - they have to satisfy the Bianchi identities
\begin{align}
0 &= \mathbf{H}_{k(ab} {\mathbf{f}^k}_{cd)} , \qquad 0 = {\mathbf{f}^a}_{k(b} {\mathbf{f}^k}_{cd)} + \mathbf{H}_{k(bc} {\mathbf{Q}_{d)}}^{ak} \nonumber \\
0 &= \mathbf{R}^{kab} \mathbf{H}_{kcd} + {\mathbf{Q}_k}^{ab}{\mathbf{f}^c}_{cd} - {\mathbf{f}^{(a}}_{k(c} {\mathbf{Q}_{d)}}^{b)k} \nonumber \\
0 &= {\mathbf{Q}_k}^{(ab} \mathbf{R}^{cd)k}, \qquad 0 = {\mathbf{Q}_k}^{(ab}{\mathbf{Q}_d}^{c)k} + {\mathbf{f}^{(a}}_{kd} \mathbf{R}^{bc)k}. \label{eq:GeneralisedFluxesBianchi}
\end{align}
Fluxes defined by 'potentials' $B$ resp. $\beta$ \eqref{eq:GeneralisedFluxesDef} fulfil these identities automatically. Without referring to these potentials, the Bianchi identities are motivated from flux compactifications \cite{Shelton2005} and Roytenberg-type algebras \cite{Blumenhagen2012}, where they encode a Jacobi identity. Later we will later discuss a (partial) interpretation of \eqref{eq:GeneralisedFluxesBianchi} in terms of Lie algebra cohomology.

\subsection{Generalised fluxes of the Poisson-Lie $\sigma$-model}
We study the generalised flux interpretation of the Poisson-Lie $\sigma$-model \eqref{eq:PoissonLieSigmaModelOriginal}
\begin{equation}
S = \frac{1}{2} \int \mathrm{d}^2 \sigma \ (g^{-1} \partial_+ g)^a \left(\frac{1}{g_0+\beta_0 + \Pi(g)}\right)_{ab} (g^{-1} \partial_- g)^b.\label{eq:PoissonLieSigmaModelGeneralisedFluxes}
\end{equation}
It is convenient to express the data $G_0$ and $B_0$ in terms of their dual objects
\begin{equation}
g_0 = (G_0 - B_0 G^{-1}_0 B_0)^{-1}, \quad \beta_0 = -(G_0+B_0)^{-1} B_0 (G_0-B_0)^{-1}. \nonumber
\end{equation}
Reading off $\beta$, we note, that it has two contributions, one from the choice of Dirac structure, $\beta_0$, and one from the (group dependent) Poisson structure $\Pi(g)$ determined by the dual Lie algebra structure ${\barf_c}^{ab}$.
\begin{equation}
\beta(g) = \beta_0 + \Pi(g)
\end{equation}
From $\beta$ and $g_0$ we can, in principle, calculate again $B_{ab}$ so that the Kalb-Ramond field is given by $B = B_{ab} (g^{-1} \mathrm{d} g)^a \wedge (g^{-1} \mathrm{d} g)^b$. The generalised fluxes for the Poisson-Lie $\sigma$-model, defined through \eqref{eq:GeneralisedFluxesDef}, are given by
\begin{align}
\mathbf{H}_{abc} &= (\mathrm{d} B)_{abc} + {f^d}_{(ab} B_{c)d} \nonumber \\
{\mathbf{f}^c}_{ab} &= {f^c}_{ab} \nonumber \\
{\mathbf{Q}_c}^{ab} &= {\barf_c}^{ab} + \beta_0^{d(a} {f^{b)}}_{dc} \nonumber \\
\mathbf{R}^{abc} &= - \beta_0^{m(a} {\barf_m}^{bc)} +
\beta_0^{(a|m} \beta_0^{|b|n} {f^{|c)}}_{mn}, \label{eq:PoissonLieGeneralisedFluxes}
\end{align}
using that $\Pi$ is a homogeneous Poisson bivector, i. e. the useful property  \eqref{eq:PoissonHomogeneousDerivative}. We see that the $\mathbf{Q}$-flux describes the 'dual geometric flux', modified to some extend by a non-vanishing $\beta_0$. This quantity is what arises as skewsymmetric part of the equations of motion\footnote{In the cases of interest to us in the following, $g_0$ will be the inverse of the Killing metric on $\mathfrak{g}$, so the symmetric part of the equations of motion of the Poisson-Lie $\sigma$-model \eqref{eq:PoissonLieSigmaModelOriginal}
\begin{equation}
\bar{j}_{+,a} \bar{j}_{-,b} \left( g_0^{m a} {f^{b}}_{cm}  + g_0^{m b} {f^{a}}_{cm} \right) = 0 
\end{equation}
vanishes. If this is not the case, it gives a non-trivial constraint on the field $g: \ \Sigma \rightarrow G$.}
\begin{equation}
\mathrm{d}\bar{j}_{c} + \frac{1}{2} {\mathbf{Q}_c}^{ab}  \bar{j}_{a} \wedge \bar{j}_{b} = 0 
\end{equation}
with
\begin{align}
\bar{j}_{\pm,a} = \pm \left( \frac{1}{g_0 \pm \beta_0 \pm \Pi} \right)_{ab} (g^{-1} \partial_\pm g)^b.
\end{align}
The $\mathbf{H}$-flux of the Poisson-Lie $\sigma$-model is the most difficult to access of the four fluxes. We try to approach it by expanding the Lagrangian in coordinates $g = \exp(x^k t_k)$ around the identity and reading of the first orders\footnote{It is straightforward but very tedious to compute higher orders and not insightful unless it reveals further structure.} of the $B$-field
\begin{equation}
\mathbf{B}_{ij} = (B_0)_{ij} + x^k \left\lbrace {f^a}_{k(i} (B_0)_{j)a} - (G_0)_{m(i|} {\barf_{k}}^{mn} (G_0)_{|j)m} - (B_0)_{m(i|} {\barf_k}^{mn} (B_0)_{|j)n} \right\rbrace + \mathcal{O}(x^2),
\end{equation}
from which we can read off the first order of the $\mathbf{H}$-flux
\begin{align}
\mathbf{H} &= \mathbf{H}^{(0)}_{ijk} + \mathcal{O}(x) \nonumber\\
\text{with} \quad \mathbf{H}^{(0)} &= {f^a}_{(ki} (B_0)_{j)a} + (B_0)_{m(i} {\barf_j}^{mn} (B_0)_{k)n} + (G_0)_{m(i} {\barf_k}^{mn} (G_0)_{j)n},
\end{align}
we see that it is quite different from what might have been expected - we even see a dependence on the (constant) metric $G_0$. 

The fluxes satisfy the Bianchi identities \eqref{eq:GeneralisedFluxesBianchi}. This is ensured by the fact, that we defined them originally in terms of potentials $\beta$ resp. $B$. So we can read of some properties of the $\mathbf{H}$-flux already, without needing to calculate it. 

\subsection{The action of the non-abelian $T$-duality group on generalised fluxes}
We study now how NATD transformations \eqref{eq:NATDgroupNew} act on Poisson Lie $\sigma$-models in term of these fluxes. The focus lies on NATD $B$- and $\beta$-shifts, because the others are not giving further insights. In particular we are interested to see, whether NATD $B$- resp. $\beta$-shifts leave indeed $\mathbf{H}$- resp. $\mathbf{R}$-flux invariant.
\begin{itemize}
\item A \textit{NATD $\beta$-shift} by a constant bivector $r$ has two effects on $\beta$ in \eqref{eq:PoissonLieSigmaModelGeneralisedFluxes}:
\begin{align}
\beta_0 \quad &\rightarrow \quad \beta_0 - r, \nonumber \\
\Pi \quad &\rightarrow \quad \tilde{\Pi}, \qquad \text{with} \quad {\overline{F}_c}^{ab} = {\barf_c}^{ab} \rightarrow {\barf_c}^{ab} + r^{m(a} {f^{b)}}_{mc}.
\end{align}
Under these transformations the background transforms as
\begin{align}
G_0 + B_0 \quad &\rightarrow \quad (G_0 + B_0) \left(r(G_0 + B_0) + \mathbb{1}\right)^{-1} \nonumber \\
{\mathbf{f}^c}_{ab} \quad &\rightarrow \quad {\mathbf{f}^c}_{ab}, \qquad {\mathbf{Q}_c}^{ab} \quad \rightarrow \quad {\mathbf{Q}_c}^{ab}\nonumber \\
\mathbf{R}^{abc} \quad &\rightarrow \quad \mathbf{R}^{abc} - \left(  r^{(a|m} r^{|b|n} {f^{|c)}}_{mn} - r^{m(a} {\barf_m}^{bc)} \right). \label{eq:PoissonLieGeneralisedFluxesBetaShift}
\end{align}
So iff $r$ fulfils the NATD group condition for $\beta$-shifts \eqref{eq:beta-shiftCondition} the $\mathbf{R}$-flux is invariant. As expected from their interpretation of being NATD $B$-shifts in the dual picture, they act as gauge transformations on the $\mathbf{R}$-flux. $\mathbf{f}$- and $\mathbf{Q}$-flux are unchanged, but $\mathbf{H}$ changes non-trivially according to a $\beta$-shift by $r$ on $G_0 + B_0$.

\item \textit{NATD $B$-shifts} by a constant $\sigma_{ab}$ act as
\begin{align}
B_0 \quad &\rightarrow \quad B_0 - \sigma, \nonumber \\
{f^c}_{ab} \quad &\rightarrow \quad {f^c}_{ab} + \sigma_{m(a} {\barf_{b)}}^{mc}
\end{align}
with an according change of $\Pi \rightarrow \tilde{\Pi}$. As mentioned already earlier the action of NATD $B$-shifts on $\sigma$-models of type \eqref{eq:PoissonLieSigmaModelGeneralisedFluxes}, although being duality symmetric to $\beta$-shifts on the bialgebra level, is more complicated here because the formulation of the $\sigma$-model (and also the generalised fluxes) is not duality symmetric.\footnote{The NATD transformations in the weaker sense of \eqref{eq:NATDgroupWeak} only act as gauge transformation on $\mathbf{R}$, as they preserve the integrability of $\mathfrak{g}^\star$ but not of $\mathfrak{g}$.} Indeed the action of $B$-shifts on the generalised fluxes is messier, as also the $\mathbf{f}$- and ${\mathbf{Q}}$-flux transform.
\begin{align}
\mathbf{H}^{(0)}_{abc} \quad &\rightarrow \quad \mathbf{H}^{(0)}_{abc} + \left(  \sigma_{(a|m} \sigma_{|b|n} {\barf_{|c)}}^{mn} - \sigma_{m(a} {f^m}^{bc)} \right) \nonumber \\
{\mathbf{f}^c}_{ab} \quad &\rightarrow \quad {\mathbf{f}^c}_{ab} + \sigma_{m(a} {\barf_{b)}}^{mc} , \label{eq:PoissonLieGeneralisedFluxesBShift}
\end{align}
which again is a gauge transformation on $\mathbf{H}^{(0)}$, iff $\sigma_{ab}$ fulfils the NATD $B$-shift condition \eqref{eq:B-shiftCondition}. We did not give the transformations of $\mathbf{Q}$- and $\mathbf{R}$-flux explicitly. They are given by the modified structure constants ${F^c}_{ab}$ and the O$(d,d)$-action of $\varphi_B$ on $\beta_0$.
\end{itemize}

\subsection{Associativity, $\mathbf{R}$-flux and Lie algebra cohomology}
In contrast to a generic non-geometric flux background, the Poisson-Lie $\sigma$-model is based on a very geometric object, the bialgebra $\mathfrak{d}$. It is logical to ask, how the bialgebra manifests itself in terms of the 'physical' fluxes ($\mathbf{H}$, $\mathbf{f}$, $\mathbf{Q}$, $\mathbf{R}$) and also how the non-geometric fluxes connect to such geometric objects, in order to understand their meaning better.

One of the key points in our cases is that the $\mathbf{f}$-flux of the Poisson-Lie $\sigma$-model represents Lie algebra structure constants. It will turn out that we can understand the $\mathbf{Q}$- and $\mathbf{R}$-flux in terms of Chevalley-Eilenburg Lie algebra cohomology on this Lie algebra. This idea was mentioned and demonstrated in some examples in \cite{Hassler2016}. For the abelian case this connection was investigated in detail in \cite{Bakas2013} - we will generalise these results here. For a brief review on Lie algebra cohomology and our conventions see appendix \ref{chap:Cohomology}.
\vspace*{0.4cm}\\
\textbf{Simple case.} Let us start with
\begin{equation}
B_0 = \beta_0 = 0
\end{equation}
in \eqref{eq:PoissonLieGeneralisedFluxes}, so that the generalised fluxes take the simple form
\begin{align}
\mathbf{H}^{(0)}_{abc} &= (G_0)_{m(a} {\barf_b}^{mn} (G_0)_{c)n}, \qquad {\mathbf{f}^c}_{ab} = {f^c}_{ab} \nonumber \\
{\mathbf{Q}_c}^{ab} &= {\barf_c}^{ab}, \qquad \mathbf{R}^{abc} =0 \label{eq:GeneralisedFluxesQDef}
\end{align}
in terms of the bialgebra structure constants. It turns out that the Bianchi identities \eqref{eq:GeneralisedFluxesBianchi} for \eqref{eq:GeneralisedFluxesQDef} encode the Jacobi identities of the bialgebra $\mathfrak{d}$ resp. $(\mathbf{f},\mathbf{Q})$. The NATD $\beta$-shift leaves ${\mathbf{Q}_c}^{ab}$ invariant.\footnote{Formally the $\mathbf{Q}$-flux is invariant under arbitrary, not only NATD, $\beta$-shifts.} This suggests that dual structures ${\barf_c}^{ab}$ to ${f^c}_{ab}$, which are related by adding a 1-coboundary, are equivalent as long the the resulting structure still fulfils the Jacobi identity. So the $\mathbf{Q}$-flux represents $\left[{\overline{f}_c}^{ab}\right] \in H^1(\mathfrak{g},\mathfrak{g}\wedge\mathfrak{g})$.

The vanishing $\mathbf{R}$-flux ensures the Jacobi identity for elements $\left[{\overline{f}_c}^{ab}\right]$. So in this simple case Poisson-Lie $\sigma$-models are classified by
\begin{equation}
{\mathbf{Q}_c}^{ab} \in H^1(\mathfrak{g},\mathfrak{g}\wedge \mathfrak{g}) \quad \text{and} \quad \mathbf{R} = 0.
\end{equation}
The $\mathbf{f}$-flux represents the structure of $\mathfrak{g}$, of course. From ($\mathbf{f}$, $\mathbf{Q}$, $\mathbf{R}$) alone, the bialgebra $\mathfrak{d}$ cannot be determined uniquely, but only up to a NATD $\beta$-shift.
\vspace*{0.4cm} \\
If we introduce a \textbf{non-vanishing $\beta_0$} to the previous setting, the $\mathbf{H}$-,$\mathbf{Q}$- and $\mathbf{R}$-fluxes are modified. Let us only comment on the changed role of the non-geometric fluxes $\mathbf{Q}$, $\mathbf{R}$. The $\mathbf{R}$-flux seems to introduce two obstruction to the previous interpretation of the $\mathbf{Q}$-flux as representing some dual structure constants to the $\mathbf{f}$-flux. The first is, what might have been expected: The trivector $\mathbf{R}$ introduces a 'non-associative' deformation on the $\mathbf{Q}$-flux Jacobi identity via its Chevalley-Eilenburg 1-coboundary
\begin{equation}
{\omega_k}^{abc} \equiv {(\delta \mathbf{R})_k}^{abc} = {\mathbf{f}^{(a}}_{ki} \mathbf{R}^{bc)i} = {\mathbf{Q}_k}^{i(a} {\mathbf{Q}_i}^{bc)} \in [0] \in H^1(\mathfrak{g},\mathfrak{g}\wedge\mathfrak{g}\wedge\mathfrak{g}).
\end{equation}
The second possible implication of the $\mathbf{R}$-flux on the $\mathbf{Q}$ could be, following the Bianchi identities \eqref{eq:GeneralisedFluxesBianchi}, that also the mixed Jacobi identity (1-cocycle condition) of the would-be bialgebra ($\mathbf{f}$,$\mathbf{Q}$) is violated:
\begin{align}
0 &= \mathbf{R}^{kab} \mathbf{H}_{kcd} + {\mathbf{Q}_k}^{ab}{\mathbf{f}^k}_{cd} - {\mathbf{f}^{(a}}_{k(c} {\mathbf{Q}_{d)}}^{b)k} \nonumber.
\end{align}
But, if we use the particular form of the generalized fluxes of the Poisson-Lie $\sigma$-model in \eqref{eq:GeneralisedFluxesDef} and the calculation \eqref{eq:B-shiftJacobiCheck}, we have independently of $\mathbf{H}$, that
\begin{equation}
{\mathbf{Q}_k}^{ab}{\mathbf{f}^k}_{cd} - {\mathbf{f}^{(a}}_{k(c} {\mathbf{Q}_{d)}}^{b)k} = 0.
\end{equation}

So, in the generic case the fluxes $\mathbf{f}$, $\mathbf{Q}$, $\mathbf{R}$ describe a kind of quasi-bialgebra structure, corresponding to a triple 
\begin{equation}
\left(\mathfrak{g}, \mathbf{Q} \in H^1(\mathfrak{g},\mathfrak{g}\wedge \mathfrak{g}), (\delta \mathbf{R}) \in H^1(\mathfrak{g},\mathfrak{g}\wedge\mathfrak{g}\wedge\mathfrak{g}) \right)
\end{equation}
generalising the bialgebra definition via 1-cocycles by allowing a non-associativity described by a trivector $\mathbf{R}$. The $\mathbf{H}$-flux is non-vanishing but 'decouples' from the other three fluxes in terms of the Bianchi identities \eqref{eq:GeneralisedFluxesBianchi}. 

The original bialgebra $\mathfrak{d}$ structure ($f$, $\barf$) and the background $\beta$-field $\beta_0$ cannot be deduced uniquly from $\mathbf{f}$, $\mathbf{Q}$ and $\mathbf{R}$, but only up to NATD $\beta$-shift as also in the simpler case.
\vspace*{0.4cm} \\
For \textbf{abelian algebras} $\mathfrak{a}$, there are no non-trivial Chevalley-Eilenburg coboundaries, so the space of $M$-valued $n$-cocycles is $H^n(\mathfrak{a},M)$. Moreover 
\begin{align}
H^1(\mathfrak{a},\mathfrak{a}\wedge\mathfrak{a}) \simeq H^2(\mathfrak{a},\mathfrak{a}) \quad \text{and} \quad H^1(\mathfrak{a},\mathfrak{a}\wedge\mathfrak{a}\wedge\mathfrak{a}) \simeq H^3(\mathfrak{a},\mathfrak{a}),
\end{align}
which reproduces the results from \cite{Bakas2013}. In constrast to the construction in this article, the Chevalley-Eilenburg cohomology groups $H^2(\mathfrak{g},\mathfrak{g})$ resp. $H^3(\mathfrak{g},\mathfrak{g})$ have been suggested to define non-commutative deformations (by a $\mathbf{Q}$-flux) resp. their non-associativity (described by an $\mathbf{R}$-flux) in \cite{Bakas2013}. 2-cocyles in $H^2(\mathfrak{g},\mathfrak{g})$ generate deformations, namely central extensions of $\mathfrak{g}$ by $\mathfrak{g}$ (treated as a vector space, becoming a central subalgebra). But this setup is not applicable to generalised geometry, as it is not possible to define an Ad-invariant O$(d,d)$-metric on these kinds of centrally extended algebras. \pagebreak

\section{Comments and applications}

\subsection{Drinfel'd doubles and generalised double field theory?}
The formulation of double field theory doubles coordinates in order to make $T$-duality manifest - the physical space is then reached after applying a constraint, the section condition (see \cite{Tseytlin1990,Tseytlin1991,Siegel1993,Hull2009,Zwiebach2011,Aldazabal2013,Hohm2013,Berman2015}). There have been approaches to incorporate non-abelian $T$-duality resp. Poisson Lie $T$-duality in double field theory \cite{Hull2009b,Reid-Edwards2010,Blumenhagen2015a,Blumenhagen2015,Hassler2016,Hassler2017}.

If we would like to make the NATD group \eqref{eq:NATDgroupNew} a manifest symmetry of a theory on a doubled space, a natural candidate for this doubled space is the Drinfel'd double $\mathcal{D}$ and a (strong) section condition then is mediated by the projection onto the (local) Dirac structures $\mathfrak{d} = \mathfrak{d}^+ \perp \mathfrak{d}^-$. There are multiple possible candidates for mathematical structures describing this 'splitting' of a Drinfel'd double, some of which could be used for constructions proposed in \cite{Vaisman2012,Vaisman2012a,Freidel2017,Svoboda2018}.

We will briefly introduce some natural candidates for such a splitting structure and argue that a para-complex structure is the most natural of these and allows us to view double field theory on Drinfel'd doubles in the framework of double field theory on para-Hermitian manifolds \cite{Vaisman2012a,Freidel2017,Svoboda2018}.

\subsubsection*{Canonical para-complex structure}
Given a bialgebra $\mathfrak{d}$ and a basis of a Manin triple decomposition $\{t_a , \bar{t}^a \}$, a canonical object describing the splitting is the linear operator
\begin{equation}
J(t_a) = t_a, \qquad J(\bar{t}^a) = - \bar{t}^a. \label{eq:ParaComplexStructure}
\end{equation}
This is an almost para-complex structure because $J^2 = \mathbb{1}$ and it has $d$-dimensional $\pm 1$-eigenbundles. $J$ is chosen in a way, that these eigenbundles are also maximally isotropic subspaces w.r.t. to $\braket{ \ | \ }$. $J$ is integrable as its Nijenhuis-tensor
\begin{equation}
N_J(X,Y) = - J^2([X,Y]) + J([J(X),Y] + [X,J(Y)]) - [J(X),J(Y)]
\end{equation}
vanishes for $X,Y \in \mathfrak{d}$. More precisely
\begin{align}
N_J(t_a , t_b) &= 0 \quad \Leftrightarrow \quad [\mathfrak{g} , \mathfrak{g}] \subset \mathfrak{g} \nonumber \\
N_J(\bar{t}^a , \bar{t}^b) &= 0 \quad \Leftrightarrow \quad [\mathfrak{g}^\star , \mathfrak{g}^\star] \subset \mathfrak{g}^\star \label{eq:ParaComplexStructureClosure} \\
N_J(t_a,\bar{t}^b) &\equiv 0 \equiv N_J(\bar{t}^a , t_b). \nonumber 
\end{align}
This opens a new perspective on $J$: Given a $2d$-dimensional Lie algebra with an Ad-invariant O$(d,d)$-metric, then the choice of a complementary pair of maximally isotropic subspaces w.r.t. to the O$(d,d)$-metric defines an almost (para-)complex structure $J$. These subspaces are closed subalgebras, iff the almost (para-)complex structure is integrable. Thus a Manin triple decomposition $(\mathfrak{d},\mathfrak{g},\mathfrak{g}^\star)$ can equivalently be described by the pair $(\mathfrak{d},J)$ with an integrable para-complex structure $J$.

The invariance group of the integrability of $J$ is exactly  the NATD group \eqref{eq:NATDgroupNew}.

\subsubsection*{Non-degenerate 2-form}
Given a metric $\braket{ \ | \ }$ and a (para)-complex structure $J$ it is possible to complete a compatible triple ($\eta$, $J$, $\omega_J$) with a non-degenerate two-form $\omega_J$ via
\begin{equation}
\omega_J(X,Y) = \braket{ J(X) | Y } . \label{eq:compatible2form}
\end{equation}
Considering the O$(d,d)$-metric and the para-complex structure $J$ \eqref{eq:ParaComplexStructure} above we get
\begin{equation}
\omega_J = t_a \wedge \bar{t}^a .
\end{equation}
With help of the Maurer-Cartan structure equation we compute
\begin{equation}
\mathrm{d} \omega = - \frac{1}{2} \left( {f^a}_{bc} t_a \wedge \bar{t}^b \wedge \bar{t}^c + {\barf_a}^{bc} \bar{t}^a \wedge t_b \wedge t_c \right),
\end{equation}
so the 2-form $\omega_J$ is symplectic, resp. $\mathcal{D}$ is a para-K\"ahler manifold, iff $\mathfrak{d}$ is abelian. For the generic case, in which we are interested in here, the apparatus for para-K\"ahler manifolds as mentioned in \cite{Vaisman2012a,Freidel2017} and thus the straightforward interpretation of the doubled space as some 'phase space' is not applicable.

\subsubsection*{Other almost (para)-complex structures}
Two other (families of) candidates for a splitting structure have been recently discussed in detail in \cite{Calvaruso2017} in a slightly different setting.\footnote{The consideration in \cite{Calvaruso2017} are more general than the one we need. They consider a Lie group, which is a semidirect product of two Lie groups of equal dimension $d$,  $\mathcal{Q} = H \ltimes K$, so the corresponding Lie algebra is $\mathfrak{q} = \mathfrak{h} \oplus \mathfrak{k}$, where $\mathfrak{h}$ is a subalgebra and $\mathfrak{k}$ is an ideal. Following this definition, they have to consider general representations $\pi: \mathfrak{h} \rightarrow \text{End}(\mathfrak{k})$ describing the action $[\mathfrak{h},\mathfrak{k}] \subset \mathfrak{k}$. The study of Drinfel'd doubles fixes the choice of representation such that it is compatible with the Ad-invariant O$(d,d)$-metric.} In the framework of Drinfel'd doubles, in which we are interested, they are only applicable for the semi-abelian Drinfel'd double $\mathcal{D} = T^\star G$ with $\mathfrak{d} = \mathfrak{g} \oplus_{\mathfrak{d}} (\mathfrak{u}(1))^d$, where we can define two almost (para)-complex structures,\footnote{In principle we could define the same structures for a generic Drinfel'd double, but only in the case of the semi-abelian double we can solve the integrability condition in a straightforward way and apply the results of \cite{Calvaruso2017}.} given a (vector space) isomorphism $\theta: \ \mathfrak{g} \rightarrow \mathfrak{g}^*=(\mathfrak{u}(1))^d$ 
\begin{itemize}
\item almost para-complex structure $\mathbb{I}: \mathfrak{d} \rightarrow \mathfrak{d}, \ (m,n) \mapsto (\theta^{-1}(n),\theta(m))$

\item almost complex structure $\mathbb{J}: \mathfrak{d} \rightarrow \mathfrak{d}, \ (m,n) \mapsto (-\theta^{-1}(n),\theta(m))$
\end{itemize}
Theorem 3.2. of \cite{Calvaruso2017}, adjusted to our case, states that these structures are integrable, iff the isomorphism $\theta$ is an 1-cocycle of $(\mathfrak{g}, \text{ad}(\mathfrak{g}) \big\vert_{\mathfrak{g}^*})$, meaning that
\begin{equation}
[m,\theta(n)] - [n,\theta(m)] - \theta([m,n]) = 0 , \quad \forall m,n \in  \mathfrak{g}. \label{eq:1CocycleCondition}
\end{equation}
There are two simple possibilities to fulfil this condition:
\begin{itemize}
\item $\mathfrak{g}$ is abelian $\Rightarrow$ $\mathfrak{d}$ is abelian. In this case any isomorphism $\theta$ will do and we could for example choose the canonical harmonic isomorphism w.r.t. to the O$(d,d)$-metric: $\sharp_\eta : \ \mathfrak{g} \rightarrow \mathfrak{g}^* , \ t_a \mapsto \bar{t}^a$, such that the integrable (para)-complex structures become
\begin{equation}
(m,n) \mapsto (\pm \flat_\eta (n),\sharp_\eta(m)).
\end{equation}

\item $\mathfrak{g}$ is a quasi-Frobenius algebra. We can define a non-degenerate 2-form $\omega = \theta_{ab} \bar{t}^a \wedge \bar{t}^b$ on $G$, where $\theta: \ t_a \mapsto \theta_{ab}\bar{t}^b$. The 1-cocycle condition means, that $\omega$ is symplectic ($\mathrm{d}\omega = 0$) resp. that $(\theta^{-1})^{ab} t_a \wedge t_b$ is a solution of the classical Yang-Baxter equation.
\end{itemize}

The canonical para-complex structure $J$ \eqref{eq:ParaComplexStructure} can be obtained via $J = \mathbb{I}\circ \mathbb{J}$. The integrability of $J$ does not depend on the integrability of $\mathbb{I}$ or $\mathbb{J}$ but only on the algebraic decomposition of $\mathfrak{d}$.

These (para)-complex structures $\mathbb{I}$ and $\mathbb{J}$ have not been applied yet to the geometric study of ordinary DFT (abelian bialgebra) or integrable deformations (quasi-Frobenius semi-abelian bialgebra case), where they might be useful. Many more details can be found in \cite{Calvaruso2017}.

\subsection{Yang-Baxter deformations as $\beta$-shifts}
\label{chap:IntDef}

The role of abelian and non-abelian resp. Poisson-Lie $T$-duality in the study of integrable deformations of string $\sigma$-models has been widely discussed. I. e. so-called $\lambda$-deformations  \cite{ Sfetsos2014,Georgiou2016,Chervonyi2016,Chervonyi2016a} were constructed as interpolations between a WZW-model and the (factorised) non-abelian $T$-dual of principal chiral model and the Yang-Baxter (also $\eta$-)deformations were introduced based on Poisson-Lie $T$-duality, generated by solutions of the modified classical Yang-Baxter equation \cite{Klimcik2002}. The integrability of latter was proven \cite{Klimcik2008,Delduc2013c}, before they were generalised to coset and supercoset $\sigma$-models, and solutions of the classical Yang-Baxter equation as generators \cite{Delduc2013c,Delduc2013b,Kawaguchi2014a}. The resulting backgrounds are (super)gravity solution if the generalising classical $r$-matrix is unimodular \cite{Borsato2016}, meaning that the resulting dual structure constants fulfil
\begin{equation}
{\barf_b}^{ab} = 0, \quad \text{for} \quad {\barf_c}^{ab} = r^{d(a} {f^{b)}}_{dc}.
\end{equation}
Nevertheless, starting from a $\kappa$-symmetric semi-symmetric type IIb supergravity background, all Yang-Baxter deformations preserve $\kappa$-symmetry and thus the resulting backgrounds are still solutions of so-called modified type IIb supergravity equations \cite{Arutyunov2015,Tseytlin2016,Borsato2016,Bakhmatov2017,Araujo2017a}. The study of many examples of homogeneous Yang-Baxter deformations  \cite{Matsumoto2014,Matsumoto2014c,Matsumoto2014d,Kawaguchi2014,Kawaguchi2014a,VanTongeren2014,Matsumoto2015,Matsumoto2015a,VanTongeren2015,VanTongeren2016,Hoare2016a, Hoare2016b,Orlando2016a} revealed that they seem to be related to $\beta$-shifts of abelian $T$-duality in case of abelian $r$-matrices. This was proven in the case of abelian $r$-matrices \cite{Osten2017} and for general $r$-matrices in case of AdS$_5\times$S$^5$ \cite{Hoare2016}. The connection of Yang-Baxter deformations and non-abelian $T$-duality became clearer in \cite{Hoare2016c,Borsato2016b,Borsato2017}. It was demonstrated, that on a purely formal level Yang-Baxter deformations are given by formal $\beta$-shifts, though there was no criterion of a connection to NATD there \cite{Sakamoto2017}. Also in case of AdS$_5\times$S$^5$-backgrounds in context of the AdS/CFT correspondence it was demonstrated that homogeneous Yang-Baxter deformations lead to Drinfel'd twists of the corresponding Hopf algebra structures on both sides of the duality \cite{VanTongeren2016,VanTongeren2017}. These aspects were investigated further in \cite{Araujo2017b,Araujo2017c}.

The non-geometric features have been already discussed in case of the (abelian) $\beta$-shifted S$^5$-background (a special case being the Lunin-Maldacena background \cite{Lunin2005}) in \cite{Frolov2005,Alday2006}, where the $\beta$-shift can be accounted for by twists of the closed string boundary conditions. The first insights into homogeneous Yang-Baxter deformations in the sense of non-geometric fluxes discussed in section \ref{chap:GeneralisedFluxes} have been found in \cite{Fernandez-Melgarejo2017}. There the $\mathbf{Q}$-flux of homogeneously Yang-Baxter deformed coset $\sigma$-models was studied in some examples and a $T$-fold interpretation of the resulting backgrounds was established.

With help of the previously developed framework of a NATD group and generalised flux analysis of the Poisson-Lie $\sigma$-model we will analyse Yang-Baxter deformed $\sigma$-models. In case of the homogeneous Yang-Baxter deformations this will proof the natural generalisation of \cite{Osten2017} for non-abelian $r$-matrices, that the notions of homogeneous Yang-Baxter deformations and NATD $\beta$-shifts of principal chiral models are exactly the same.

\pagebreak
\subsubsection{Yang-Baxter deformations}
Consider the Yang-Baxter deformed Lagrangian
\begin{equation}
S = \frac{1}{2} \int \mathrm{d}^2 \sigma \quad (g^{-1} \partial_+ g)^a \kappa_{ac} {\left( \frac{1}{\mathbb{1} - \eta R_g} \right)^c}_b (g^{-1} \partial_- g)^b, \label{eq:YB-deformedLag}
\end{equation}
where $\kappa_{ab}$ is the Killing form on the Lie algebra $\mathfrak{g}$ and $R_g = \text{Ad}_g^{-1} \circ R \circ \text{Ad}_g$ (and also $R$) is a solution to the (modified) classical Yang-Baxter equation\footnote{Let us emphasise for clarities sake the distinction between $R$-operator, related to $\beta$ by $\beta^{ab} = {R_g^a}_c \kappa^{cb}$, and the $\mathbf{R}$-flux, defined in terms of $\beta$, as $\mathbf{R} = [\beta,\beta]_S$.}
\begin{equation}
[R(t_a),R(t_b)] - R([R(t_a),t_b)+[t_a,R(t_b)]) = - c^2 [t_a,t_b].
\end{equation}
For $c=0$ the corresponding deformations are usually called (homogeneous) Yang-Baxter deformations and for $c=i$ $\eta$-deformations.

We can express $R_g$ conveniently in the language of the previous sections
\begin{equation}
{(R_g)^a}_c \kappa^{cb} = r^{ab} - \Pi^{ab}(g),
\end{equation}
where $\Pi(g)$ is the homogeneous Poisson structure corresponding to the $R$-bracket of $R$. We see that \eqref{eq:YB-deformedLag} is of the form of our above definition of a NATD $\beta$-shift starting from a principal chiral model. In order for the Yang-Baxter deformation to be a NATD $\beta$-shift, $r^{ab}$ has to fulfil \eqref{eq:beta-shiftCondition}, which for ${\barf_c}^{ab} = 0$ in case of the principal chiral model is
\begin{equation}
r^{(a|m} r^{|b|n} {f^{|c)}}_{mn} = 0,
\end{equation}
which is exactly the (homogeneous) classical Yang-Baxter equation for a bivector $r$ on $\mathfrak{g}$. This proofs the conjecture, that homogeneous Yang-Baxter deformations are exactly the same as NATD $\beta$-shifts of principal chiral models.

Let us also discuss the Yang-Baxter deformed model at the level of generalised fluxes.
\begin{align}
{\mathbf{Q}_c}^{ab} &= 0 \nonumber \\
\mathbf{R}^{abc} &= - \eta^2 c^2 \kappa^{ak} {f_k}^{bc}, \label{eq:YB-deformedFluxes}
\end{align}
The $\mathbf{Q}$-flux clearly vanishes because the Yang-Baxter deformation is a formal $\beta$-shift. The $\mathbf{R}$-flux of the deformed model vanishes as expected for $c=0$. For $c=i$, the $\eta$-deformation, we see that  \eqref{eq:YB-deformedLag} is a realisation of an $\mathbf{R}$-flux background and of course the $\eta$-deformation is not a NATD $\beta$-shift.

Let us compare these results with the ones in \cite{Fernandez-Melgarejo2017}, where the authors studied examples of deformed coset $\sigma$-models. The key results there was, that the deformed backgrounds should be interpreted as $T$-folds, because going around closed cycles we pick up a monodromy in $\beta$, which is described by a non-vanishing $\mathbf{Q}$-flux. So our result \eqref{eq:YB-deformedFluxes} of vanishing $\mathbf{Q}$-flux might seem overly simplistic, but in constrast to Yang-Baxter deformed principal chiral models we have $\beta^{ab} = {\left(R_g \circ P\right)^a}_c \kappa^{cb}$ for Yang-Baxter deformed coset $\sigma$-models, where $P$ is the projector on the coset algebra. This projector makes the algebraic situation much more diverse, which appearently also leads to a non-vanishing $\mathbf{Q}$-flux.

\subsubsection{Bi-Yang-Baxter deformations}
A very natural generalisation of the Yang-Baxter deformation \eqref{eq:YB-deformedLag}, which turns out to be still integrable \cite{Klimcik2014,Delduc2016}, is
\begin{equation}
S = \frac{1}{2} \int \mathrm{d}^2\sigma (g^{-1} \partial_+ g)^a \kappa_{ac} {\left( \frac{1}{\mathbb{1} - \xi R - \eta R_g} \right)^c}_b (g^{-1} \partial_- g)^b. \label{eq:BiYB-deformedLag}
\end{equation}
Originally it was introduced for $R$ being a solution mcYBe($i$), but is also integrable for $R$ solution of classical Yang-Baxter deformation.\footnote{Formally it corresponds then to separate $\beta$-shifts on the isometries $G_R \times G_L$ for the principal chiral model with separate scales $\xi$ and $\eta$. This is becomes clear as 
\begin{align*}
(g^{-1} \partial_+ g)^a \kappa_{ac} {\left( \frac{1}{\mathbb{1} - \eta R_g} \right)^c}_b (g^{-1} \partial_- g)^b = ( \partial_+ g \ g^{-1})^a \kappa_{ac} {\left( \frac{1}{\mathbb{1} - \eta R} \right)^c}_b (\partial_- g \ g^{-1} )^b.
\end{align*}}
For our purpose we generalise to the case, where the $R$-bracket fulfils the Jacobi identity, and which is not generically integrable.

Let us rewrite for our purposes
\begin{equation}
\beta^{ab} = {(\xi R + \eta R_g)^a}_c \kappa^{cb} = (\xi + \eta) r^{ab} - \eta \Pi^{ab}(g),
\end{equation}
The generalised fluxes for this model take the form
\begin{align}
{\mathbf{Q}_c}^{ab} &= - \xi {f^{(a}}_{cd} r^{b)d} \nonumber \\
\mathbf{R}^{abc} &= - (\xi+\eta)^2 r^{(a|m} r^{|b|n} {f^{|c)}}_{mn}.
\end{align}
For $\xi = - \eta$ they become especially simple
\begin{align}
\mathbf{H}_{abc}^{(0)} &= \eta \kappa_{am} \kappa_{bn} {f^{(m}}_{cd} r^{n)d} = \eta \overline{f}_{abc} \nonumber \\
{\mathbf{f}^c}_{ab} &= {f^c}_{ab} \nonumber\\
{\mathbf{Q}_c}^{ab} &= \eta{f^{(a}}_{cd} r^{b)d} = \eta {\overline{f}_c}^{ab} \nonumber \\
\mathbf{R}^{abc} &= 0 .
\end{align}
Then \eqref{eq:BiYB-deformedLag} describes an $\mathbf{R}$-flux free model which is (in case $R$ is not a solution to the classical Yang-Baxter equation) not related to the principal chiral model via a NATD $\beta$-shift, because the $\mathbf{Q}$-flux is changed. 

\section{Conclusion}
In this paper we looked rather schematically at Poisson-Lie $\sigma$-model \eqref{eq:PoissonLieSigmaModelOriginal} from different perspectives. These perspectives included methods from generalised geometry and double field theory, non-geometric fluxes, integrable deformations and Lie algebra cohomology. 

After setting the stage in introducing bialgebras and Poisson-Lie $T$-duality, which motivated the duality group
\begin{align*}
\text{NATD group}(\mathfrak{d}) = \{ \text{Manin triple decompositions of } \mathfrak{d} \}
\end{align*}
of a Poisson-Lie $\sigma$-model corresponding to the bialgebra $\mathfrak{d}$, we proposed a method to get some insights into this group, which is some subgroup of O$(d,d)$. We studied the conditions of the typical O$(d,d)$-transformations, i.e. factorised dualities, GL-transformations, $B$-shifts and $\beta$-shifts such that they lie in this non-abelian $T$-duality group, and found in some simple cases an explicit interpretation of these transformation on the (non-doubled) Lagrangian level. The analysis revealed some interesting structures, but also that the above definition does not give the whole (classical) duality structure of a Poisson-Lie $\sigma$-model: the standard non-abelian $T$-duality of a principal chiral model w. r. t. a subgroup, turns out \textit{not} to correspond to a Manin triple decomposition of $\mathfrak{d}$. Based on this we also proposed a slightly generalised version of the non-abelian $T$-duality group:
\begin{equation}
\text{mod. NATD group} \left(\mathfrak{d}\right) = \left\lbrace \text{Manin pair decompositions of } \mathfrak{d} \right\rbrace, \nonumber
\end{equation}
which would also allow for other, slightly more general basic objects than a bialgebra.

The analysis, which was done in this paper, was using aspects of generalised geometry and the notion of generalised fluxes. We computed the fluxes explicitly and it turns out that the non-geometric fluxes $\mathbf{Q}$ and $\mathbf{R}$ have some meaning in terms of Lie algebra cohomology on $\mathfrak{g}$. In general ($\mathbf{f}$, $\mathbf{Q}$, $\mathbf{R}$) describe a kind of quasi-bialgebra, where the $\mathbf{R}$-flux describes the failure of dual structure constants $\mathbf{Q}$ to associate. The Poisson-Lie $\sigma$-model in general and the integrable $\eta$-deformed principal chiral model in particular seem to be realisations of a \textit{classical} string in a non-associative $\mathbf{R}$-flux background.

In the last section we studied the Yang-Baxter deformations with help of this non-abelian $T$-duality group and generalised fluxes. This proved that homogeneous Yang-Baxter deformations are nothing else than non-abelian $T$-duality $\beta$-transformations following our definition.

\subsection*{Outlook}

Regarding the non-abelian $T$-duality group, there are some open questions. The extensions to geometries with spectator coordinates, coset geometries, supergroups, etc. seem straightforward. A more difficile is to understand the web of dual models: already in the simple example of the principal chiral model to a group $\mathcal{G}$, this web of dualities might become very complicated, depending on the amount of 2-cocycles and solutions to the classical Yang-Baxter equation on $\mathfrak{g}$. Another issue is, what a rigorous motivation and the structure of a non-abelian $T$-duality group in the spirit of the modified definition \eqref{eq:NATDgroupWeak} is.

The approach in this paper, which is based on the (undoubled) classical string $\sigma$-model \eqref{eq:PoissonLieSigmaModelOriginal} seems to yield different results than the approach for a Poisson-Lie $T$-duality invariant DFT in \cite{Hassler2016,Hassler2017}. For example in our approach a potential $\beta$-field is a very natural object, which would allow for a non-trivial $\mathbf{R}$-flux as its Jacobiator, whereas in \cite{Hassler2016,Hassler2017} it does not seem possible so far to find solutions to the section condition, which result in an $\mathbf{R}$-flux background. It would be very interesting to implement the backgrounds corresponding to the Poisson-Lie $\sigma$-models in the framework of \cite{Hassler2016,Hassler2017} or, if this does not work, to understand what causes the difference between the two approaches. One reason could be, that in \cite{Hassler2016,Hassler2017} the Dirac structures are defined globally on the Drinfel'd double $\mathcal{D}$, whereas for the construction \cite{Klimcik1995} of Poisson-Lie $T$-duality, that we use in section \ref{chap:Review}, it is sufficient to define a (local) decomposition of the bialgebra $\mathfrak{d}$.

The connection of deformed classically integrable $\sigma$-models to generalised geometry, which was one subject of this paper, might help to understand the nature of non-geometric backgrounds better, but also to approach hidden symmetries in integrable $\sigma$-models using these algebraic structures. Moreover this connection could shed light into the proof of equivalence between $\lambda$- and $\eta$-deformations via Poisson-Lie $T$-duality and analytic continuation, or a suitable generalisation to superspace could give new perspectives on the connection between modified type IIb supergravity and $\kappa$-symmetry.

\section*{Acknowledgements}
We thank Ben Hoare, Konstantinos Sfetsos and Stijn van Tongeren for interesting discussions during this work, Henk Bart for comments on the draft and Falk Hassler and Kentaroh Yoshida for pointing out issues in the first version. The work of D.L. is supported by the ERC Advanced Grant No. 320045 'Strings and Gravity' and the Excellence Cluster Universe.

\pagebreak
\appendix
\section{Lie algebra cohomology}
\label{chap:Cohomology}
For convenience of the reader we state the basic notions and the results used in this paper of Chevalley-Eilenburg Lie algebra cohomology here (see e.g. \cite{Chevalley1948,Kosmann2004}). 
\subsection{Definition}
A $M$-valued $k$-\textit{cochain} $u$ is a $k$-linear skewsymmetric map $u: \ \mathfrak{g} \wedge ... \wedge \mathfrak{g} \rightarrow M$, being an element of $\Lambda^k \mathfrak{g}^\star \otimes M$. In general we take $M$ to be a vector space of a representation $\rho$ of $\mathfrak{g}$ and as such a $\mathfrak{g}$-module.

We can define a coboundary $\delta u \in \Lambda^{k+1}\mathfrak{g}^\star \otimes M$ of such a $k$-cochain via a \textit{coboundary operator} $\delta$, defined as
\begin{align}
\delta: \ \Lambda^k \mathfrak{g}^\star \otimes M &\rightarrow \Lambda^{k+1} \mathfrak{g}^\star \otimes M, \ u \mapsto \delta u \nonumber \\
\delta u (x_0,x_1,...,x_k) &= \sum_{i=0}^k (-1)^i \rho(x_i).\left( u(x_0,...,\hat{x}_i,...,x_n)\right) \label{eq:Coboudary}\\
&{} \quad + \sum_{i<j} (-1)^{i+j} u([x_i,x_j],x_0,...,\hat{x}_i,...,\hat{x}_j,...,x_k). \nonumber 
\end{align}
As $\delta^2 = 0$, we can define the \textit{$k$-th cohomology vector space} of $\mathfrak{g}$ w.r.t. the representation $\rho$ as usual as
\begin{equation}
H^k(\mathfrak{g},M) = \frac{\text{$k$-cocycles}}{\text{$k$-coboundaries}} .
\end{equation}

\subsection{Results}
\subsubsection*{Compact Lie algebras and lower cohomology groups}
Let $\mathfrak{g}$ be a compact semi-simple (finite-dimensional) Lie algebra, then the Chevalley-Eilenburg cohomologies $H^1(\mathfrak{g},M)$ and $H^2(\mathfrak{g},M)$ are trivial \cite{Lichnerowicz1988}. From this two statements follow:
\begin{itemize}
\item There are no non-abelian compact quasi-Frobenius algebras.
\item There are no non-trivial solutions of the classical Yang-Baxter equation \eqref{eq:CYBE} on compact Lie algebras, except on their abelian subalgebras.
\end{itemize}
On the other hand in the \textit{non-compact} case we can hope to find non-trivial elements of $H^1$ and $H^2$. For a nice discussion of the corresponding ($M$-valued) group cohomology see appendix of \cite{Bakas2013}.

\subsubsection*{$H^2(\mathfrak{g},M)$ and central extensions}
Given a Lie algebra $\mathfrak{g}$ we consider the exact sequence
\begin{equation}
0 \hookrightarrow \mathfrak{h} \overset{i}{\hookrightarrow} \mathfrak{e} \overset{s}{\twoheadrightarrow} \mathfrak{g} \twoheadrightarrow 0, \label{eq:CentralExtensionSequence}
\end{equation}
so $\text{Ker}(s)=\text{Im}(i)$ is an ideal of the extended algebra $\mathfrak{e}$ and $\mathfrak{g}$ can be reproduced by $\mathfrak{g} \simeq \frac{\mathfrak{e}}{\text{Im}(i)}$. If $\mathfrak{h}$ is abelian, $\mathfrak{e}$ is called \textit{central extension}. 

$H^2(\mathfrak{g},M)$ has the nice interpretation as central extensions of $\mathfrak{g}$ by the $\mathfrak{g}$-module $M$, then considered to be an abelian algebra. Let us demonstrate the isomorphism explicitly for the trivial representation\footnote{The extension to higher dimensional modules $M$, in case they are vector spaces themselves, corresponds basically to independent central extensions of the form for each generator of $M$ as discussed below.}  $M = \mathbb{F}$, which is the case of interest in section \ref{chap:NATDsubgroups}.
\begin{itemize}
\item Given a 2-cocycle $\omega$ on $\mathfrak{g}$, let us define a bracket $[ \ , \ ]^\prime$ on the extended algebra which viewed as a $\mathbb{F}$-vector space is $\mathfrak{e} = \mathfrak{g} \oplus \mathbb{F}Z$, where $Z$ is the generator of $\mathfrak{h}$
\begin{align}
[m_1 + r_1 Z, m_2 + r_2 Z]^\prime &=  [m_1, m_2] + \omega(m_1,m_2) Z, \label{eq:CentralExtensionCocycle1}\qquad \forall m,n \in \mathfrak{g}, \ r_1,r_2 \in \mathbb{F}.
\end{align}
The 2-cocycle condition on $\omega$ is equivalent to $[ \ , \ ]^\prime$ fulfilling the Jacobi identity. Also, if we take $\mathfrak{s}$ in \eqref{eq:CentralExtensionSequence} to be the canonical projection of $[ \ , \ ]^\prime$ back on $\mathfrak{g}$, this naturally reproduces the original Lie bracket $[ \ , \ ]$ on $\mathfrak{g}$. So $\mathfrak{s}$ is a Lie algebra homomorphism.

\item Given a central extension $\mathbb{F} \overset{i}{\hookrightarrow} \mathfrak{e} \overset{s}{\twoheadrightarrow} \mathfrak{g}$, consider the diagram:

\begin{center}
  \begin{tikzcd}
    \mathfrak{g}\times\mathfrak{g} \arrow{dr}{\epsilon} \arrow[swap]{d}{\omega} & & \\
    \mathbb{F} \arrow{r}{i} & \mathfrak{e} \arrow[rightharpoonup,shift left=0.5ex]{r}{s} \arrow[swap,leftharpoondown,shift right=0.5]{r}{l} & \mathfrak{g}  
  \end{tikzcd}
\end{center}
where $l$ is a section of $s: \ \mathfrak{e} \rightarrow \mathfrak{g}$, i.e. $s\circ l = \text{id}_\mathfrak{g}$. We use $l$ to define the map
\begin{equation}
\epsilon: \mathfrak{g}\times\mathfrak{g}\rightarrow \mathfrak{e}, \quad \epsilon (m_1,m_2) = l([m_1,m_2]) - [l(m_1),l(m_2)] \ \text{for}  \ m_1,m_2 \in\mathfrak{g}, \label{eq:CentralExtensionSequence2}
\end{equation}
for which holds $\epsilon(m_1,[m_2,m_3]) + \ \text{c.p} = 0$ due to Jacobi identities in $\mathfrak{e}$ and $\mathfrak{g}$. Thus $\epsilon$ and $\omega = i^{-1} \circ \epsilon$ are $\mathfrak{e}$- resp. $\mathbb{F}$-valued 2-cocyles on $\mathfrak{g}$. This proves that for any central extension we can find a 2-cocycle.
\end{itemize}

\section{Yang-Baxter equations and bialgebras}
\label{chap:Rbrackets}

\subsection{Classical Yang-Baxter equation and Poisson-Lie groups}
Given a Lie group $\mathcal{G}$, we can define a Poisson bracket compatible with group multiplication \cite{Drinfeld1983} on $C^\infty(\mathcal{G})$ as
\begin{equation}
\{ f , g \} \equiv \Pi^{ab} X_a[f] X_b[g], \label{eq:PoissonBracketGroup}
\end{equation}
where $X_a \in \chi(\mathcal{G})$ are the left (right) invariant vector fields associated with the $t_a \in \mathfrak{g}$. Evaluated at the identity ($\Pi^{ab}(e) = r^{ab}$) skew-symmetry and Jacobi-identity of $\{ \ , \ \}$ are equivalent to 
\begin{equation}
r = r^{ab} t_a \otimes t_b \ \in \ \mathfrak{g} \otimes \mathfrak{g}
\end{equation}
being skew-symmetric ($r^{ab} = - r^{ba}$) and fulfilling the classical Yang-Baxter equation (cYBe)
\begin{equation}
[r_{12},r_{13}] + [r_{12},r_{23}] + [r_{13},r_{23}] \equiv [r , r]_S = 0, \label{eq:CYBE}
\end{equation}
where the indices are tensor space indices.

Such a Poisson structure on $\mathcal{G}$ induces a Lie algebra structure on $\mathfrak{g}^*$, the dual vector space of the Lie algebra  $\mathfrak{g}$, generated by  $d$ dual generators $\bar{t}^a$:
\begin{align}
[ \bar{t}^a , \bar{t}^b] &\equiv \left[ \mathrm{d} X^a\big|_e , \mathrm{d} X^b\big|_e \right] := \mathrm{d} \{ X^a , X^b \}\big|_e = \left( \partial_c \Pi^{ab} \right) \mathrm{d} X^c\big|_e \equiv {\overline{f}_c}^{ab} \bar{t}^c  .\label{eq:DualStructureConst}
\end{align}
The Jacobi identity for a Lie algebra with structure constants ${\overline{f}_c}^{ab}$ is fulfilled due to $r$ fulfilling the cYBe, if ${\overline{f}_c}^{ab}$ is given in terms of $r^{ab}$ and the original structure constants by
\begin{align*}
{\overline{f}_c}^{ab} = r^{ad} {f^b}_{dc} - r^{bd} {f^a}_{dc},
\end{align*}
which can be shown by expanding the Poisson bivector
\begin{equation}
\Pi^{ab} = r^{ab} - r^{(a|d} {f^{|b)}}_{dc} X^c + ... \ ,
\end{equation}
for coordinates $X^a$ associated to Lie algebra generators $t_a$. Viewed from the Lie group of the dual Lie algebra $\mathfrak{g}^\star$, $\Pi^{ab}$ is a closed two-form. 

With such a pair of structure constants ${f^c}_{ab}$, ${\overline{f}_c}^{ab}$ on $\mathfrak{g}$ resp. $\mathfrak{g}^\star$, we can define a Lie bialgebra via \eqref{eq:Bialgebra}, where the Jacobi identity \eqref{eq:BialgebraJacobi} is fulfilled, if $r$ is a solution of the classical Yang-Baxter equation.

\subsection{Complex double}
Consider the complexification of a (real) simple Lie algebra $\mathfrak{g}$ 
\begin{equation}
\mathfrak{g}^\mathbb{C} := \mathfrak{g} \otimes_\mathbb{R} \mathbb{C}
\end{equation}
with the corresponding involution $\tau(m \otimes u) = m \otimes \bar{u}, \ \forall m \in \mathfrak{g}, \ u \in \mathbb{C}$ fixing $\mathfrak{g} \subset \mathfrak{g}^\mathbb{C}$. Suppose we have a solution of the \textit{non-split} mcYBe $R$, so that $\mathfrak{g}$ with the corresponding $R$-bracket $[ \ , \ ]_R$, denoted by $\mathfrak{g}_R$, is a Lie algebra and thus $R^- = R - i: \ \mathfrak{g}_R \hookrightarrow \mathfrak{g}^\mathbb{C}$ is injective and Lie algebra homomorphism due to  \eqref{eq:mCYBEhomomorphism}. The sequence\footnote{This is \textit{not} an exact sequence.}
\begin{equation}
\mathfrak{g}_R \overset{R^+}{\hookrightarrow} \mathfrak{g}^\mathbb{C} \overset{\mathcal{I}}{\twoheadrightarrow} \mathfrak{g}
\end{equation}
with $\mathcal{I}$ being the projection on the imaginary part w.r.t. $\tau$ and $\mathcal{I}\big\vert_{\text{Im}(R^-)}$ being bijective describes the 'splitting' of $\mathfrak{g}^\mathbb{C}$ into a Drinfel'd double
\begin{equation}
\mathfrak{g}^\mathbb{C} = \mathfrak{g} \oplus_\mathfrak{d} \mathfrak{g}_R \equiv \text{Ker}(\mathcal{I}) \oplus_\mathfrak{d} \text{Im}(R^-) \label{eq:ComplexDouble}
\end{equation}
w.r.t. to the bilinear form
\begin{equation}
\langle m + in \vert m^\prime + i n^\prime \rangle :=  \gamma\left( m , n^\prime \right) + \gamma\left( n , m^\prime\right), \quad \text{for} \ m,m^\prime,n,n^\prime \in \mathfrak{g},
\end{equation}
where $\gamma$ is the (non-degenerate) Killing form of $\mathfrak{g}$ and $R$ has to be skewsymmetric for $\langle \ \vert \ \rangle$ to be non-degenerate and $g_R$ to be Lagrangian.

\textbf{Example.} Take the Lie algebra $\mathfrak{g}^\mathbb{C} = \mathfrak{sl}(2,\mathbb{C})$ with standard generators $\{ h , e , f \}$ fulfilling the commutation relations
\begin{align*}
[h,e] &= e, \quad [h,f] = -f, \quad [e,f] = 2 h
\end{align*}
and with Killing form $\gamma(h,h) = \frac{1}{2} \gamma(e,f) = 1$. Then
\begin{equation}
\mathfrak{g}^\mathbb{C} = \mathfrak{sl}(2,\mathbb{C}) = \text{span}_{\mathbb{R}}\left\{i h,i(e + f),e-f \right\} \oplus_\mathbb{d} \text{span}_{\mathbb{R}}\left\{h,e,ie \right\}
\end{equation}
is the decomposition into the compact real form $\mathfrak{su}(2)$ and the borel algebra $\mathfrak{b}$ of positive roots and is of the form \eqref{eq:ComplexDouble} for the (canonical) Drinfel'd-Jimbo $R$-operator
\begin{equation}
R: \ h \mapsto 0, \quad e \mapsto ie, \quad f \mapsto -i f .
\end{equation}
The above structure w.r.t. to the Cartan-Weyl basis can be generalised in a straightforward manner. For reviews see \cite{Hoare2017,Vicedo2015}.

\textbf{Group decomposition.} In contrast to the Poisson-Lie case the decomposition can be defined also at the level of the corresponding Lie group $G^\mathbb{C}$:
\begin{equation}
G^\mathbb{C} = G G_R = G_R G,
\end{equation}
where $G$ resp. $G_R$ is the Lie group to $\mathfrak{g}$ resp. $\mathfrak{g}_R$, meaning that we can write each $g \in G^\mathbb{C}$ as $g = h_1 h_2$ for $h_1 \in G$ and $h_2 \in G_R$ and vice versa. For the canonical Drinfel'd-Jimbo $R$-operator on a compact Lie algebra as above this is equivalent to the Iwasawa decomposition. In the case where the real form is non-compact we can find such a decomposition for each non-connected component.

\subsection{Real double}
Consider a (real) simple Lie algebra $\mathfrak{g}$, admitting a solution $R$ of the split mcYBe and the direct (Lie algebra) sum of $\mathfrak{g}$ with itself:
\begin{align}
\mathfrak{d} = \mathfrak{g} \oplus \mathfrak{g}
\end{align}
with projectors $p_{1,2}: \ \mathfrak{d} \rightarrow \mathfrak{g}$ on the first/second copy and the diagonal subalgebra 
\begin{align}
\mathfrak{g}^\delta := \left\{(m,m)|m \in \mathfrak{g} \right\} \subset \mathfrak{d}.
\end{align}
$R^\pm = (R \pm 1)$ are Lie algebra homomorphisms between $\mathfrak{g}_R$ and $\mathfrak{g}$ similarly to the complex double case. Consider the maps
\begin{align*}
\iota: \ &{} \mathfrak{g}_R \hookrightarrow \mathfrak{d}, \ m \mapsto (R^+(m) , R^-(m)) \\
\nu: \ &{} \mathfrak{d}=\mathfrak{g} \oplus \mathfrak{g} \twoheadrightarrow \mathfrak{g}, \ (m,n) \mapsto \frac{1}{2}(m-n)
\end{align*}
which are injective respectively surjective Lie algebra homomorphisms, such that the sequence
\begin{align}
\mathfrak{g}_R \overset{\iota}{\hookrightarrow} \mathfrak{d} \overset{\nu}{\twoheadrightarrow} \mathfrak{g}
\end{align}
describes a splitting of $\mathfrak{d}$, as $\nu\vert_{\text{Im}(\iota)}$ is bijective. The decomposition
\begin{align}
\mathfrak{d} = g^\delta \oplus_\mathfrak{d} g_R \equiv \text{Ker}(\nu) \oplus_\mathfrak{d} \text{Im}(\iota)
\end{align}
is a Drinfel'd double w.r.t. to the bilinear form on $\mathfrak{g} \oplus \mathfrak{g}$
\begin{equation}
\langle(m,n) \vert (m^\prime , n^\prime) \rangle := \gamma(m,n^\prime) - \gamma(n,m^\prime), \quad \text{for} \ m,m^\prime,n,n^\prime \in \mathfrak{g}.
\end{equation}
Again the non-degeneracy of $\langle \ \vert \ \rangle$ and also Lagrangian property of $\mathfrak{g}_R$ depends on the skew-symmetry of $R$.

As in the complex case, there is also a decomposition (for each connection component) $G \otimes G \simeq G^\delta G = G G^\delta$. 

\textbf{Example.} Consider the 'analytic continuation' $R \rightarrow -iR$ of the Drinfel'd-Jimbo $R$-operator of the non-split mcYBe
\begin{equation}
R: \ h \mapsto 0, \quad e \mapsto  e, \quad f \mapsto -f .
\end{equation}
for the above generators of $\mathfrak{sl}(2,\mathbb{C})$. Of course it is now not an endomorphism on the compact real form $\mathfrak{su}(2)$ anymore, but on the split real form $\mathfrak{sl}(2,\mathbb{R})$. The $R$-bracket then is
\begin{equation}
[h,e]_R = e, \quad [h,f]_R = 0,\quad [e,f]_R = 0,
\end{equation}
so, similar to the complex double case $\mathfrak{g}_R \simeq \mathfrak{b} = \text{span}_\mathbb{R}(h,e,ie)$. We can decompose the as
\begin{align}
\mathfrak{d} &= \mathfrak{sl}(2,\mathbb{R}) \oplus \mathfrak{sl}(2,\mathbb{R}) = (\mathfrak{sl}(2,\mathbb{R}))^\delta \oplus_{\mathfrak{d}} \mathfrak{g}_R \nonumber \\
&= \mathfrak{span}_\mathbb{R} \left( h \oplus h, e \oplus e, f \oplus f \right) \oplus_\mathfrak{d} \mathfrak{span}_\mathbb{R} \left( h \oplus (-h), e \oplus 0 , 0 \oplus f \right).
\end{align}
This again directly generalises to general split real forms and their doubles.

\subsection{Lie bialgebras without Yang-Baxter equations?}
\label{chap:WithoutYB}
In the physics literature mostly bialgebras corresponding to solutions of Yang-Baxter equations were considered. These are bialgebras, where the defining 1-cocycle is a 1-coboundary of a 0-cocycle $r = r^{ab} t_a \wedge t_b$, so \eqref{eq:Bialgebra1cocycle1} is satisfied automatically, and where the (modified) classical Yang-Baxter equation \eqref{eq:mCYBE} for the operator ${R^a}_b = r^{ac} \kappa_{cb}$, with the Killing form $\kappa$ on $\mathfrak{g}$ holds, which is a sufficient condition for \eqref{eq:Rbracket} to fulfil \eqref{eq:Bialgebra1cocycle2}.

So these span only a subspace of 1-coboundary bialgebras - in general $R$ does not have to be a solution of the (modified) classical Yang-Baxter equation. Also there are non-coboundary 1-cocycles fulfilling \eqref{eq:Bialgebra1cocycle2}. Let us demonstrate now that these more general bialgebra structures are not at all exotic. 

\subsubsection*{bialgebras on the torus.}
The trivial example are possible bialgebras to an abelian Lie algebra. There are no non-trivial Chevalley-Eilenburg coboundaries on an abelian algebra, so any structure constants ${\barf_a}^{bc}$ correspond to a 1-cocycle on an abelian algebra, fulfil the Jacobi identity and resultantly define a possible bialgebra, the semi-abelian bialgebra $\left(\mathfrak{u}(1)\right)^{d} \oplus_{\mathfrak{d}} \mathfrak{g}^\star$.

\subsubsection*{bialgebras for $\mathfrak{sl}(2,\mathbb{R})$}
Consider the Lie algebra $\mathfrak{sl}(2,\mathbb{R})$ with generators $(h,e,f)$ and the commutation relations
\begin{equation}
[h,e] = e, \quad [h,f] = -f \quad \text{and} \quad [e,f] = 2h.
\end{equation}
There are well-known $r$-matrices on $\mathfrak{sl}(2,\mathbb{R})$:
\begin{itemize}
\item the jordanian $r$-matrix $r = h \wedge e$, solving the classical Yang-Baxter equation.
\item the Drinfel'd-Jimbo $r$-matrix $r = c \ e\wedge f$ being a solution to the modified classical Yang-Baxter equation.
\end{itemize}
As such they correspond to bialgebra structures. But simply solving conditions \eqref{eq:Bialgebra1cocycle1} and \eqref{eq:Bialgebra1cocycle2} shows, that also a generic skewsymmetric $r$-matrix $r = r^{ab} t_a \wedge t_b$ generates coboundary bialgebras. Up to a total scale which can be absorbed into the definition of the dual generators the most general $r$-matrix leading to an 1-coboundary satisfying \eqref{eq:Bialgebra1cocycle2} is\footnote{It seems to be a three parameter space of bialgebras, but we use the fact that the $r$-matrix
\begin{equation}
r = e\wedge f + \frac{1}{2} \left(A^2 h \wedge e + \frac{1}{A^2} h \wedge f \right)
\end{equation}
is equivalent to the jordanian $r$-matrix $r=h\wedge a$ via conjugation of the $\mathfrak{sl}(2)$-generators
\begin{equation}
t^\prime = StS^{-1} \quad \text{with} \quad S = \left( \begin{array}{cc} - \frac{\sqrt{2}}{A} & 0 \\ - \frac{1}{\sqrt{2} A} & - \frac{ A}{\sqrt{2}} \end{array} \right).
\end{equation}}
\begin{equation}
r = A e\wedge f + B h \wedge e,
\end{equation}
which is simply the sum of the jordanian and the Drinfel'd-Jimbo $r$-matrix. There seem to be no non-coboundary bialgebra structures on $\mathfrak{sl}(2,\mathbb{R})$.

\subsubsection*{bialgebras for $\text{AdS}_3$}
$\text{AdS}_3$ has the isometry algebra $\mathfrak{sl}(2,\mathbb{R}) \oplus \mathfrak{sl}(2,\mathbb{R})$. There is a big amount of solutions already from Yang-Baxter type solution (e.g. abelian \cite{Osten2017}, jordanian \cite{Hoare2016}, Drinfel'd-Jimbo,...). but other 1-coboundary and 1-cocycle Lie algebras exist. Let us give a non-Yang-Baxter 1-coboundary, which is closely related to the Drinfel'd-Jimbo solution of the mcYBe:
\begin{equation}
r = A\left( e_1 \wedge f_1 + \frac{1}{2} h_1 \wedge h_2 \right) + B \left( e_2 \wedge f_2 - \frac{1}{2} h_1 \wedge h_2 \right).
\end{equation}
Let us note, that there are only unimodular dual Lie algebras for $\text{AdS}_3$, which are 1-coboundaries of solutions to the classical Yang-Baxter equation.

\bibliographystyle{jhep}
\bibliography{References}

\end{document}